\documentclass[12pt, a4paper, twoside, openright]{book}

\usepackage{vuwthesis} 

\usepackage{palatino} 

\usepackage{url} 
\usepackage{graphicx}
\usepackage{caption}
\usepackage{amsmath}
\usepackage{mathrsfs}
\usepackage{amssymb}
\usepackage{xcolor}
\usepackage{soul}
\usepackage{subcaption}
\usepackage{adjustbox}
\def\Ak{\mathbb{A}}
\def\ak{\alpha_k}
\def\en{\epsilon_0}
\def\B#1{\mathbf{#1}}
\def\ddt#1{\frac{\partial #1}{\partial t}}
\def\bfrac#1#2{\left(\frac{ #1}{#2}\right)}
\def\DimD#1{\hat{\nabla}_{#1}}
\def\ex#1{\times 10^{#1}}

\def\eten{\times 10^{10}}

\def\dg{^{\circ}}

\begin{document}

\frontmatter


\title{Mixed Flow Ionospheric Aerodynamics}
\author{Faun Watson}

\subject{Physics}
\abstract{
The increasingly congested near earth environment requires accurate orbital modelling to prevent collision events that threaten access to space infrastructure. Ionospheric aerodynamics are the largest non-conservative source of orbital perturbations. Work done by Capon et al \cite{CaponBible} provided a prediction method for drag forces on charged bodies in low earth orbit based on applying approximations of single species flows to representative plasma simulations. Recent work on mixed species plasma sheaths \cite{MixedSheathKnewIt}\cite{ThreeSpeciesPlasmaSheath} motivates investigation of this assumption.\\
Mixed species plasma flows over charged bodies with representative LEO conditions were simulated, focusing on orbital motion limited O+ dominated flows, and sheath limited H+ dominated flows. The significance of small concentrations of different ion species was analysed in terms of aerodynamic effects on the body. Introduction of small concentrations of H+ into a flow dominated by O+ molecules showed significant disruptions to the wake structure, coupled with an increase in drag coefficient for a satellite in that flow. Introducing heavier O+ molecules into a sheath dominated H+ flow found little disruption to the wake structure.\\ 
A multi-species drag model was proposed, building on work in \cite{CaponBible}. This model was found to qualitatively describe changes in wake structure, but consistently over predicted drag coefficients. This finding suggests that the influence of mixed species flows on ionospheric aerodynamics involves significant contributions from inter-species interactions not described adequately by considering the separate species as either separate flows, or one flow with adjusted parameters.
}

\mscthesisonly


\maketitle

\chapter*{Acknowledgments}\label{C:ack}

First and Foremost I would like to acknowledge my supervisors, Dr Tulasi
Parashar and Dr Jakub Glowacki, for their relentless support through this
project. Your encouragement and support has allowed me to grow as an
academic, and meant the world to me. Without the countless hours of
debugging, and insistence to ”come back with that written down”, this
work would not have been possible
I would like to also acknowledge the help of Dr. Chris Capon, for pro-
viding the pdFoam code, as well as numerous insights as to the nature of
these simulations. Without this help, I would still be puzzling over the
oddities of OpenFOAM syntax. Similarly, thanks goes to Chris Acheson,
for the sanity found in commiserating together over the many riddles of
OpenFOAM.

\tableofcontents


\mainmatter



\chapter{Introduction}\label{C:intro}
Satellite infrastructure in low earth orbit (LEO) between 200  and 2000km comprises the majority of active satellites \cite{UCSSatellites}, with these satellites playing a critical role in tasks from weather monitoring and communications, to the research performed by the Hubble telescope and international space station.Operating at LEO offers a number of advantages over higher orbits, most notably that the low altitudes enable a relatively low launch cost, as well as lower signal latency, and require less powerful instruments to image and communicate with the planet's surface. These advantages come with the cost that the low earth environment is the most crowded, with over 27,000 objects large enough to be tracked\ \cite{UCSSatellites}. Orbits in this range must also account for non-negligible atmospheric drag forces.\par
In 2021, two SpaceX satellites had near misses with the Chinese space station in LEO, passing within 4km after the space station maneuvered to a safe height \cite{AlJazeera}. While this collision was successfully avoided by this maneuvering, the potential for future collisions grows daily with the increased debris in LEO. The potential hazards of such interactions are illustrated clearly in the 2009 collision between the Iridium-33 communications satellite and defunct Russian military satellite Kosmos-2251, in which a direct collision resulted in the complete destruction of both satellites, and over 1000 pieces of debris $>$10cm being released into orbit. With the speed of such satellites and debris typically greater than  5km/s , the kinetic energy in any such debris pieces holds the potential to cripple or destroy any satellites with which it may collide, placing even more fragmented debris on orbital trajectories \cite{debrismitigationRex}.\\

 Orbital debris models in \cite{Kessler} suggested that orbital space debris would become a significant problem during the next century, unless operational procedures and launch constraints are changed appropriately. The 'Kessler Syndrome' predicted in that seminal work described an environment too saturated with high speed debris for it to be possible to reasonably avoid any collision events, leading to all bodies at similar orbits being similarly fragmented, and making this orbital band both unusable to orbiting bodies and unsafe to pass through in transit to higher orbits \cite{Kessler}.\\
 A key part of mitigating orbital debris and thus preventing these predictions coming to pass is the maneuvering of active satellites to avoid 'conjunction events' with other orbiting bodies. This requires both up to date knowledge and control of a satellites position, and the ability to predict trajectories of any debris approaching it. As these collision avoidance maneuvers are typically not quick to perform, this often requires ground observers to predict potential conjunction paths multiple orbital periods in advance, requiring relatively high accuracy \cite{satelliteManeuvering}.\par

The force acting on a small (relative to Earth), unpowered orbiting satellite are typically dominated by conservative forces (in this case, Earth's gravity) with the remaining force given by \cite{FuenteDrag}:
\begin{equation}
    \B{F_{NC}}=\B{F}_{rad}+\B{F}_{thermal}+\B{F}_{Aero}
\end{equation}
Where $\B{F}_{rad}$ is the radiation pressure on the satellite (from both solar and terrestrial sources), $\B{F}_{thermal}$ is the force from a thermal radiation imbalance caused by uneven temperature distributions across the body, and $\B{F}_{Aero}$ is aerodynamic forces. In the LEO environment, the largest of these nonconservative forces is $\B{F}_{Aero}$ \cite{FuenteDrag}. Associated with this is the largest uncertainties, with errors accumulating as the result of assumptions regarding drag coefficients, gas-surface interactions \cite{MoeMoe}, and atmospheric wind and weather \cite{DoornbosDensityModel}\cite{MSISE00}.\par
Assuming the dominant aerodynamic force acts parallel to the motion of the satellite (ie. drag forces dominate over any wind or induced lift forces), the acceleration of a satellite can be written as:
\begin{equation}
    a_{Drag}=-\frac{1}{2}\frac{C_{D}A_{\perp}}{m_B}\rho_mv^2_B
    \label{DragEquation}
\end{equation}
 Where $A_{\perp}$ is the cross sectional area of the body with respect to the flow of gas,  $m_B, v_B$ are the mass and speed of the body, $\rho_m$ is the mass density of the flow, and $C_{D}$ is the neutral drag coefficient. This equation is common in aerodynamics\cite{WingTheoryAbbot}, and functions on the assumption that most drag is proportional to the incident kinetic energy of the flow, with the proportion given by the dimensionless drag coefficient\footnote{It is important to note that the drag coefficient is NOT the drag force, but a measure of drag experienced relative to flow conditions. The Eiffel tower ($C_D=\approx2.0$) has a comparable drag coefficient to the neutral satellite approximation ($C_D\approx2.2$) in their respective environments, but trying to interchange one with the other would be unwise.} 
 $C_{D}$. \\
 Historically, a common set of assumptions based on work by Cook et al. \cite{CookDragCoefficient} resulted in a drag coefficient of 2.2 being used as a routine estimation for simple satellites in LEO. This drag estimation remained largely in place until the 21st century, when a variety of works (\cite{FuenteDrag}\cite{ISSCharging}\cite{Hastings30V}\cite{OlivieraWeather}\cite{OlsenPurvisCharging}) called these assumptions into question. One notable early example is the work by \cite{MoeMoe}, in which it was demonstrated that the drag coefficient of a satellite could vary by up to 40$\%$ between simple geometries, with significant uncertainties introduced by the use of different models for interaction between particles and the satellite surface . Vallado and Finkleman \cite{ValladoAndFinkleman} presented an assesment of a range of drag models on orbit propagation, and proposed the use of a standardised set of parameters for drag modelling.\\
previous atmospheric density models have often been derived using satellite acceleration data, using this $C_D=2.2$ assumption. Doornbos and Finkleman \cite{DoornbosDensityModel} compared two thermospheric density models, both incorporating measurements from the ERS-12 satellite. Acceleration measurements of the ERS-12 were combined with Eqn\eqref{DragEquation} to infer the mass density $\rho_m$ in the path of the satellite. Comparing the two density models showed frequent areas of significant ($>5\%$) discrepancies between the two models in various regions, concluded by Doornbos and Finkleman to be the result of "imperfect correlation between space weather proxies and densities" (In this case, the proxy in question is the satellite deceleration).\\
This selection of studies agree in their conclusions that the $C_D=2.2$ neutral drag approximation was not sufficient to explain variations in drag forces experienced by orbiting bodies, nor the significant increase in $C_D$ for bodies passing through a geomagnetic storm. \\
It therefore becomes prudent to consider potential sources of these discrepancies. One such source, proposed as a solution by Capon et al, was charged ionospheric aerodynamics, the distinct aerodynamics associated with a charged environment \cite{Caponthesis}. 
\par
Aerodynamic interactions between orbiting bodies and the ionosphere (a charge environment) is fundamentally different to atmospheric interactions in a neutral flow, such as in the thermosphere. When an object is immersed in a plasma, it acquires a floating potential ($\phi_B$) with respect to the freestream plasma based on the sum of currents into or out of the surface. In LEO, the plasma environment contains electrons with thermal velocities orders of magnitude higher than the orbital velocity of the orbiting body, which is itself higher than the ion thermal velocity ($v_{e,T}>>v_B>>v_{i,T}$, known as a mesothermal flow). This velocity distribution results in a negative floating potential developing on the orbiting body \cite{ISSCharging}, and a region of charge discontinuity, known as the 'plasma sheath'. In this region, ions are accelerated towards the body while electrons are repelled, resulting in an area of relatively low ion density, the thickness of which depends on the floating potential, as well as the ion distribution of the flow \cite{BenilovLangmuirSheath}. \\
\begin{figure}[h!]
         \centering
     
         \begin{minipage}[c]{\textwidth}
             \centering
             \includegraphics[width=0.99\textwidth]{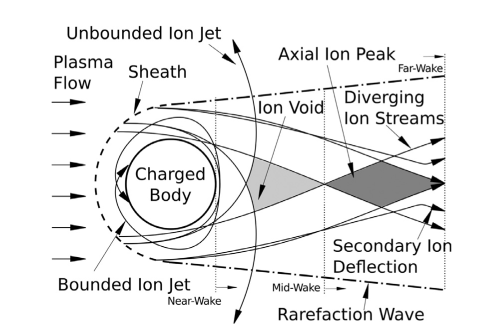}
              \centering
         \end{minipage}
        \captionsetup{width=0.7\linewidth}
         \caption{Schematic of LEO plasma-body interaction features, showing various deflected ion paths emerging from the sheath boundary. (reproduced from Capon et al \cite{CCScaling} with permission from author) }
         \label{CCWakeDiagram}
     \end{figure}
The structure of this plasma sheath, elaborated on more fully in Chapter \ref{C:wakeStructures}, is crucial to the understanding of ionospheric aerodynamic interactions \cite{CaponBible}, with several major distinctions from a traditional neutral flow. Most notably, the plasma sheath provides a larger effective collection area, with ions being accelerated towards the forebody that would otherwise not be in its path, and would not have contributed to drag forces (Fig.\ref{CCWakeDiagram}. Ions along the plasma sheath edges will also be deflected into the wake by the body charge. These deflections are usually a purely electromagnetic interaction, with an indirect momentum exchange through electrostatic attraction to the body (see Section \ref{section:SingleSpeciesflows} for more details and exceptions). This means that when predicting charged aerodynamic drags, it is important to accommodate for both direct and indirect aerodynamic interactions, rather than just the traditional direct interactions.\par
Despite the fact that these charged aerodynamics are known to increase drag forces beyond those in a neutral interaction, they remained unmodelled in the neutral density inferences above, due to the conclusions of Cook \cite{CookDragCoefficient}. Cook's drag coefficient relied on a series of assumptions by Brundin et al \cite{BrundinChargedDrag}, primarily:
\begin{itemize}
    \item The maximum ion density of the flow is approximately 10$\%$ of the neutral number density.
    \item Singly ionised oxygen is the dominant source of any charged aerodynamic force (O+).
    \item The surface potential of LEO objects is never more negative than -0.75V with respect to the quasi-neutral freestream plasma.
\end{itemize}
Based on which it was predicted that the ionospheric aerodynamics would be dominated by neutral interactions. With the growth of power electronics, and novel equipment on satellites, surface potentials can reach -30V \cite{Hastings30V}, while improved measurement and modelling shows some regions can temporarily hold charges in excess of -500V from interaction with solar weather in polar orbits \cite{ISSCharging}. \par
While the study of LEO plasma interactions is not new, previous work focused on either calculating ion and electron surface currents (\cite{Whipple81}\cite{GODDLaFramboiseCurrentCollection}\cite{Stone1981SpaceAerodynamics}), or focusing on wake phenomena in the context of charging (\cite{Stone1981SpaceAerodynamics}\cite{Hastings30V}). Charging and ion currents are intrinsically intertwined with the aerodynamics of these objects, but a complete understanding involves analysis of the momentum exchange involved in these interactions. \\

Work done at UNSW Canberra \cite{CaponBible} modelled charged interactions using physics based particle in cell (PIC) codes to simulate single species plasma flows over charged bodies. In this work, plasma flow simulations and experiments were parametrised and used to form a response surface to predict $C_D$ for various flow conditions in LEO. Applying this empirical rule to a range of LEO orbits predicted orbits above 450km would frequently experience an increase of at least 10$\%$ total force around an orbit from charged drag forces, with charged drag in sections of orbit spiking up to 40$\%$ of total drag. This analysis poses a number of questions around the accuracy of inferred density atmospheric models, as well as providing a convenient framework for further work on ionospheric aerodynamics.\par
\begin{figure}[h!]
         \centering
         \begin{minipage}[c]{0.9\textwidth}
             \centering
             \includegraphics[width=0.96\textwidth]{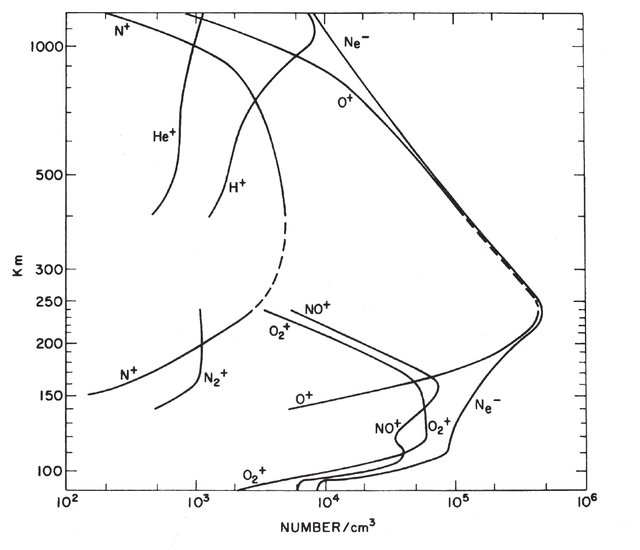}
              \centering
         \end{minipage}
        \captionsetup{width=0.7\linewidth}
         \caption{Plot of ion densities as a function of height for the lowest 1000km of LEO, showing O+ as the dominant source of ions for the range 200-800km. Neutral atoms are excluded from plot.Of particular interest to this thesis is the region of 800-1000km, in which H+ and O+ have similar ion densities.(Adapted from Johnson, 1969, fig 1 with permission)}
         \label{IonosphereComposition}
     \end{figure}

Flows simulated for the empirical response surface developed in \cite{CaponMain} were exclusively single species, and the resulting conclusions rely on the assumptions that a section of an orbits path is either dominated by a single species of plasma (usually H+ or O+), or that the behaviour of the mixed flow is consistent with that of a single species of ion. Neutral species behaviour at these altitudes is assumed to still behave according to the cannonball assumption(modelled as a simple flow with a drag coefficient of 2.2).
The former assumption is true for a large range of LEO altitudes, with most LEO plasma flows being dominated by either H+ or O+ ions, below or above 900km, respectively. (Fig.\ref{IonosphereComposition}). This leaves a range of ~150km on either side with significant ($>10\%$) contributions of charge from the less dominant species. Other Ion species are also present at these altitudes in lower numbers, leading to a flow containing a mixture of various ions, either dominated by O+ or the much lighter H+ species. While the latter assumption is not explicitly tested in \cite{CaponBible}, the $C_D$ of various plasma flows is shown to be very sensitive to small variations in wake structure, and the mixed species cases shown in \cite{CCScaling} showed significantly altered wake structures, indicating the inclusion of a species of significantly different behaviour (specifically the charge-mass ratio) will likely provide a non-negligible contribution to total drag 
. Work performed by Severn et al \cite{MixedSheathKnewIt} in the form of laser flourescence measurements of a plasma sheath found the introduction of lighter He+ ion species to an existing Ar+ flow resulted in the Ar+ ions entering the sheath with an increased velocity, though the mechanism behind this was not explained in this work. This therefore motivates analysis of the behaviour of this set of mixed flows, and analysis of the relevance of the single species approximation to conclusions made in \cite{CaponBible}.
\section{Research Questions and Aim}

The background material laid out above motivates an investigation of the nature of the assumptions employed in Capon et al (\cite{CaponBible}). This work therefore aims to address the research question\\
How does the presence of multiple species of plasma affect the ability to predict ionospheric aerodynamic forces?\\
This offers a number of avenues of study, including analysis of known satellites orbits, scaling parameter analysis, grain based plasma physics, and many others. For simplicity, the specific goals of this analysis are confined to:
\begin{enumerate}
    \item Extend set of parameters in Capon et al \cite{CCScaling} to account for multi-species interactions in the formulation of a control surface.
    \item Simulation of multi-species flow using particle in cell code developed at University of New South Wales (UNSW) Canberra, and subsequent drag analysis.
\end{enumerate}
In this way, the extensive parametric study of single species flows performed in \cite{Caponthesis} can be directly compared to mixed plasma flows, thereby allowing mixed species analysis to be an extension, rather than replacement, of the existing response surface model.\par
In studying mixed species plasma flows, the difference in interactions of mixed species compared to single species flows can be largely attributed to either differing charge-mass ratios\cite{FORTOVDustyPlasmas}, differing gas-surface interactions \cite{MoeMoe}, or collisional chemical interactions\cite{BirdBookDSMC}. In the plasmas being modelled (mesothermal LEO flows), the mean free path is on the order of kilometres \cite{LiuMeanFreePath}, and therefore the analysis of chemical reactions is neglected in this work. Numerous gas-surface interaction models have been employed in previous works, and found to contribute significant variation between different simulated drag coefficients \cite{MoeMoe}. While this provides an exciting avenue for future research, this work holds a constant model of diffuse, neutralised reflection (elaborated on in Chapter \ref{C:pdFoam}), in order to allow direct comparison of results to those in \cite{CaponBible}. The primary point of analysis will therefore be the mixing of ions with varying charge:mass ratios. For simplicity, analysis will be limited to mixes of only two species, chosen to be singly ionised H+ and O+. The choice of these species has 3 distinct motivators:
\begin{itemize}
    \item O+ and H+ represent the two extremes of charge to mass ratios in singly ionised particles in LEO.
    \item O+ and H+ are the two most populous plasma species in the LEO regions under consideration (see Fig.\ref{IonosphereComposition}).
    \item Use of O+ and H+ allow direct comparisons to the flows analysed in Capon et al \cite{CaponMain}, as these are all single species H+ or O+ flows.
\end{itemize}
Using these species, the parameters analysed were the ratio of oxygen to hydrogen ions ($\rho_{H+}:\rho_{O+}$), and the sensitivity of these results to variations in satellite charge. Specifically, analysis began with a representative O+ single species flow, and varied the ratio $\rho_{H+}:\rho_{O+}$ in steps, providing a series of simulated flows with a range of H+ and O+ concentrations. The flows were varied by either number density or mass density, investigating the effects of both parameters on drag distributions. These two series of simulations were then repeated using a lower body potential $\phi_0$, testing sensitivity of results to variations in body charge.

\chapter{PIC Simulation methods}\label{C:pdFoam}
All simulation code  used in this study was built in the OpenFoam framework \cite{CAPONPdFoam}. OpenFOAM is a computational fluid dynamics (CFD) toolbox written in C++, containing a range of solvers for partial differential equations.
The hybrid particle in cell, direct simulation monte carlo modelling code (PIC-DSMC) used for this work was developed by UNSW Canberra \cite{CAPONPdFoam}. The development and validation of this pdFoam code is covered in \cite{CaponBible}, while this section only aims to outline the basic working principles of particle in cell and DSMC modelling as used in this code, as well as modifications made to it.\par
\subsection{Choosing a plasma dynamics solver}
When choosing computational fluid dynamics (CFD) modelling tools, an important consideration is the density regimes being modelled, for which we consult the Knudsen number. The Knudsen number is given as the ratio of the mean free path length $\lambda$ of a molecule to a characteristic length scale (in this case, the satellite radius $r_0$ ). 
\begin{equation}
    K_n=\frac{\lambda}{r_0} 
    \label{Knudsen number}
\end{equation}
A low Knudsen number implies the motion of a fluid around a body is driven by frequent collisional interactions, allowing it to be modelled as a continuum flow. In this flow, the fluid is assumed to be a continous mass, and described primarily through conservation equations. A high Knudsen number implies the collisional interactions within a fluid are much less frequent, and this flow is instead described by kinetic theory. In kinetic theory, flows macroscopic quantities are calculated with statistical analysis of microscopic behaviours. Of particular interest to this study are Monte-Carlo methods, in which a system's final state is modelled by taking a large ensemble of potential states of the system and allowing these to evolve, with the error in final solution approaching zero as the number of samples increases.  The system of a small satellite in LEO is a high Knudsen number regime, with $r_0$ typically being less than a metre, and $\lambda>$1km \cite{LiuMeanFreePath}, and therefore lends itself to a kinetic theory approach. \par

The near earth environment is a collection of positively charged ions, negatively charged electrons, and neutral atoms and molecules. To describe this system, we define the phase space distribution function $f$ of species $k$ within volume element $dx_1dx_2dx_3$ as $f_k(\bf{x},\bf{c_\alpha},t)$, where $\textbf{c}_k$ and $\textbf{x}$ are the particle velocity and position, respectively, at time t. 
The evolution of $f_k$ with $t$ is described by the Boltzmann equation (\cite{maxwelStressTensor}):
\begin{equation}
    \frac{\partial f_k}{\partial t} + \underbrace{c_k\cdot\nabla_x f_k}_{Diffusion} + \underbrace{\frac{\bf{F}_k}{m_k}\cdot \nabla_c f_k}_{External Force} = \left(\frac{\partial f_k}{\partial t}\right)_{coll} 
    \label{Boltzmann equation}
\end{equation}
where the terms, from left to right, represent: the rate of change of $f_k$ with time, the diffusion component of $f_k$, and the influences of external forces $\bf{F}_k$ acting on $f_k$, with the right hand term describing the rate of change of $f_k$ as a result of particle collisions.
In a plasma, Eqn.\ref{Boltzmann equation} describes the interactions of particles of mass $m_k$ and charge $q_k$ through mutual electric and magnetic fields via the Lorentz force. In the context of LEO plasma-body interactions, the interaction may be considered electrostatic and unmagnetised  \cite{Whipple81},\cite{Stone1981SpaceAerodynamics}, allowing Maxwells equations to reduce down to Poissons equation for the electric potential.
\begin{equation}
    \bf{E}=-\nabla\phi,\quad \nabla^2\phi=-\frac{\rho_c}{\epsilon_0} 
    \label{Poisson equation}
\end{equation}
where $\epsilon_0$ is the permittivity of free space, and $\rho_c$ is the macroscopic charge density
\begin{equation}
    \rho_c=\sum_k q_k \int f_k d\bf{c}_k=\sum_k q_k n_k
    \label{Charge density}
\end{equation}
For practical systems, with multiple reacting species, and a lack of external forces, exact numerical solutions of the Boltzmann equations are unfeasible. DSMC and PIC simulations instead solve the Boltzmann method using probabilistic monte carlo methods.

\section{DSMC and PIC simulations}

The DSMC method is a technique for simulating high Knudsen number flows, where the mean free path length of particles is sufficiently high for particle motion and collisions to be decoupled over sufficiently small time step $\Delta t$.
The fundamental mechanic of DSMC is the superparticle method, which simulates the motion of particles directly, by grouping a given number of particles into a 'macroparticle', and then tracking the motion and collisions of these macroparticles. 
As the ratio of real particles to simulated particles (a parameter referred to as $N_{Equiv}$ in this work) approaches 1 (each macroparticle representing one real particle), the simulation approaches a direct calculation of the position of each of the particles present, providing a computationally intense, but mathematically intuitive, approach to simulations. 
The general DSMC algorithm used in the dsmcFoam code implemented in OpenFOAM is as follows
\begin{enumerate}
    \item \textbf{Initialisation} - The mesh is generated using blockmesh utility from parameters given in blockmeshDict file, and a number of macroparticles is assigned to each cell in the mesh. This number is given by $\frac{V_c\cdot \rho_k}{N_{Equiv}}$ for cell volume $V_c$ and species density $\rho_k$ . These macroparticles are assigned representative temperatures and velocities from the initial flow conditions supplied, modified by the Maxwell-Boltzmann distribution, producing a collection of thermalised particles with the desired average macroscopic quantities. Each Cell is also assigned the list of macroparticles it contains.
    \item \textbf{Particle Push} - Iterating across all macroparticles, the macroparticles positions are updated ballistically: $\bf{x}_k\rightarrow\bf{x}_k+\bf{v}_k\Delta t$
    \item \textbf{Update Cell Occupancy} - Each Cells list of occupying particles is updated to reflect the changed positions in step 2. 
    \item \textbf{Collision partner selection} - Within each cell, the average number of expected collisions is determined by iterating across the particles present, and (Using the No Time Counter (NTC) numerical method laid out in \cite{BirdBookDSMC}), the average number of collisions expected for the present particles over a timestep $\delta t$ is determined, and an appropriate number of pairs of particles is selected to collide.
    \item \textbf{Collisions} - the particles in a collision pair have their internal and kinetic energy redistributed according to kinetic theory described in \cite{BirdBookDSMC}, with new velocity determined by relative kinetic energies, and new directions of travel randomised (while conserving total collision momentum).
    \item \textbf{Reactions} - If particle reactions are being used, these are resolved according to rules dictated by the type of reaction. In the cases studied here, the only reaction was a neutralisation reaction upon contact with charged surface, so the particles were reflected diffusely, and their charge was changed from +1 to 0.
    \item \textbf{Update Cell occupancy} - Each cell's list of occupying particles is once again updated, this time to include changes in particle species from reaction interactions. The algorithm then returns to step 2.
\end{enumerate}
 Most of this algorithm is very intuitive, with the exception of the collisions method (known as No Time counter (NTC)). In this method, the collisions occuring within a cell are not necessarily based on the actual relative trajectories of particles, but rather the relative scattering cross section $\sigma$ of two particles. In this way, macroscopic collision properties are preserved, without the necessity of individually calculating the relative trajectories of every particle with every other, a computationally imposing task.\\

 \subsubsection{The Particle In Cell (PIC) method}
 In plasma physics, a 'strongly coupled' system is one in which the density of plasma is low enough that electric field at a point is dominated only by the closest ions, leading to a high variation between points. A weakly coupled system is one in which the plasma density is higher, and the electric field at any point is far more stable, and has a weaker coupling to the location of nearby ions. In order to describe these systems, we make use of the plasma coupling parameter $\Lambda$. For a box with side length of the Debye length (length at which a potential disturbance is screened by 1/$e$ in the plasma), the number of particles present inside is 
 \begin{equation}
    N_D=n\lambda^3_D 
    \label{DebyeBox}
\end{equation}
Where $n$ is the number density of the plasma. A system is weakly couple for large $N_D$ and strongly coupled for small $N_D$. For two ions separated by distance $a$, the electrostatic potential energy is given by 
\begin{equation}
    E_{pot}=\frac{q^2}{4\pi\epsilon_0 a} 
    \label{EPot}
\end{equation}
and from kinetic theory we also know the kinetic energy to be 
\begin{equation}
    E_k=kT
    \label{Ek}
\end{equation}
We can then combine Eqn.\ref{EPot} and Eqn.\ref{Ek} to give the plasma coupling parameter
\begin{equation}
    \Lambda=\frac{E_k}{E_{pot}}=\frac{4\pi\epsilon_0akT}{q^2} 
    \label{PlasmaParam1}
\end{equation}
Using the definition of the Debye length as $\lambda_D=(\epsilon_0kT/ne^2)$, and average interparticle distance $a=n^{-\frac{1}{3}}$ gives
\begin{equation}
    \Lambda= \frac{4\pi\epsilon_0kT}{q^2a^\frac{1}{3}}=4\pi N^\frac{2}{3}_D
    \label{PlasmaParam2}
\end{equation}
the plasma parameter is a measure of the relative kinetic and potential energy, with a Debye box containing many particles being dominated by thermal energy, and therefore a large $\Lambda$, and a sparsely populated Debye box is dominated by potential energy, for a smaller $\Lambda$. 

\begin{figure}[h!]
         \centering
         \begin{minipage}[c]{0.65\textwidth}
             \centering
             \includegraphics[width=\textwidth]{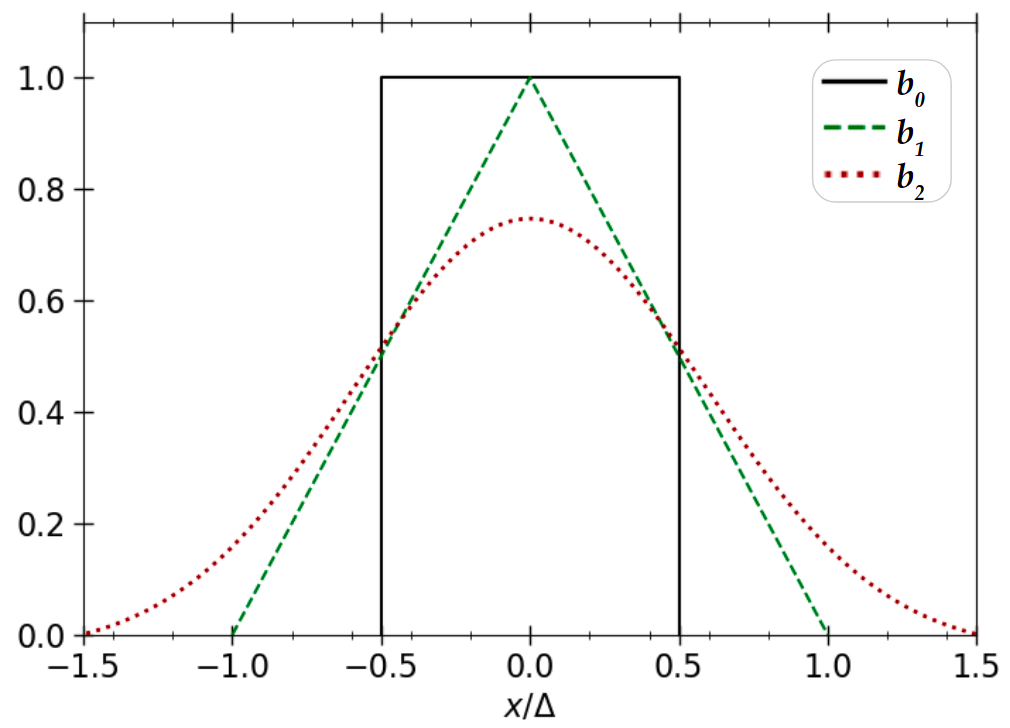}
              \centering
         \end{minipage}
        \captionsetup{width=0.91\linewidth}
         \caption{First 3 spline functions, with width given in terms of grid spacing, showing (from top) the top hat ($b_0$), wedge ($b_1$), and bell curve ($b_2$). }
         \label{BSplines}
         \end{figure}

 The particle in cell method is used to determine solutions to the Vlasov maxwell system for weakly coupled systems (small $\Lambda$). The key idea in simulation of weakly coupled systems is to not use single particles but collective clouds of them, where each computational particle represents a group of particles and can be visualised as a small piece of phase space. The computational particles are of finite size, and interact more weakly than point particles. Unlike point particles, where the force between two particles grows as they approach each other, the finite size particles behave as point particles until their surfaces overlap, where the overlap area is neutralised, and does not contribute to the force. This allows the correct plasma parameter to be achieved using fewer particles than a physical system. It also bears many similarities to DSMC modelling, with the direct simulation of particle motion through macroparticles.
 \\
The pdFoam code being used in this work is a hybrid code, combining dsmcFoam collisional methods with the PIC approach to electrostatics, and a boltzmann electron Fluid model to account for the movement of electrons in the plasma.
Two important concepts here are the shape and interpolation functions, where the shape functions $S_x$ and $S_y$ are used to define the distribution function $f_s$ within a superparticle, 
\begin{equation}
    f_p=N_{Equiv}S_x(x-x_p(t))S_v(v-v_p(t))
    \label{functional dependence on shape function}
\end{equation}
The shape function obeys certain properties:
\begin{itemize}
\item Shape functions are compact, describing a small portion of phase space (0 outside a small range)
\item the integral across any coordinate is unity
\item Shape functions are symmetric across any coordinate
\end{itemize}
The standard particle shape in $v$ coordinates is a Dirac Delta function, as a small distribution in $v$ space lets particles remain closely grouped in space during subsequent evolution. The shape function in physical space is typically a b spline, one of a series of consecutively higher order functions obtained from integrating the others. First among these ($b_0$) is a top hat function , followed by a triangle ($b_1$), and a bell curve ($b_2$) (Fig.\ref{BSplines}).

The interpolation function $W$ is the convolution of the shape function with the top hat ($b_0$) function. 
\begin{equation}
    W(x_i - x_p)=\int S_x(x-x_p)b_0\left(\frac{x-x_i}{\Delta x}\right)
    \label{interp function definition}
\end{equation}
The interpolation function allows direct computation of the cell density through summations, without the need for integrations (demonstrated in Eqn.\ref{ChargeDensityAssignment}).

The general algorithm for pdFoam simulations is given as :
\begin{enumerate}
    \item \textbf{Initialisation} - Much like DSMCFoam, a mesh is created, and Macroparticles are assigned to each cell, each particle having position $x_p$ and velocity $v_p$, and representing $N_{Equiv}$ physical particles.
    \item \textbf{Particle Push} - The equations of motion for the particles are advanced one timestep using the leapfrog scheme. (new position and velocities are calculated with an offset of $\frac{1}{2}$ a timestep, to account for the change in electric field as particle position is changed.)
    \begin{equation}
\begin{aligned}
x_p^{n+1}=x_p^n+\Delta tv_p^{n+1/2} \\
v_p^{n+3/2}=v_p^{n+1/2}+\Delta t \frac{q_s}{m_s}E_p^{n+1}
    \label{LeapFrogPush}
\end{aligned}
\end{equation}
    \item \textbf{Update Cell Occupancy} - The list of particles within a cell is updated to account for changed particle positions, as in DSMC.
    \item \textbf{DSMC Collisions} - the DSMC algorithm is used to resolve any collisions or reactions, as described previously.
    \item \textbf{Charge assignment}The charge densities in each cell are computed using 
    \begin{equation}
    \rho_i=\sum_p \frac{q_p}{\Delta x}W(x_i-x_p)
    \label{ChargeDensityAssignment}
    \end{equation}

    \item \textbf{Field computation} - The poisson equation is solved:
    
\begin{equation}
    \epsilon_0 \frac{\varphi_{i+1}-2\varphi_i+\varphi_{i-1}}{\Delta x^2}=-\rho_i
    \label{PoissonSolve}
\end{equation}
and the electric field in each cell is computed:
\begin{equation}
    E_i = -\frac{\varphi_{i+1}-\varphi_{i-1}}{2\Delta x}
    \label{EFieldCalc}
\end{equation}
    \item \textbf{Lorentz Force} - The field in the cells is used to calculate the field acting on each particle: 
    \begin{equation}
    E_p^{n+1}=\sum_i E_i W(x_i-x_p^{n+1})
    \label{E on particle laputa}
\end{equation}
which is used in the next cycle. 
    \item The algorithm then returns to step 2, and the cycle continues with the updated $E_p$, $x_p$, and $v_p$ values.
\end{enumerate}
.
\begin{figure}[!htb]
             \centering
             \includegraphics[height=0.7\textheight]{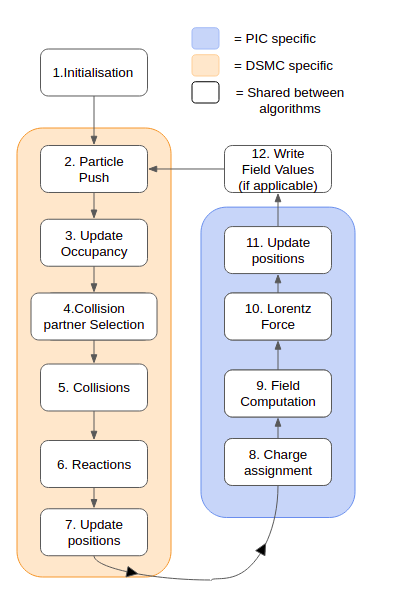}
        \captionsetup{width=0.81\linewidth}
         \caption{pdFoam algorithm summary, showing components belonging to DSMCFoam, PIC and algorithm independent components separately.}
         \label{FlowChart}
         \end{figure}

\newpage
\subsubsection{Boltzmann electron fluid}
Directly simulating electrons comes at a significant computational cost. To maintain numerical stability, $\Delta t$ must be smaller than the fastest plasma frequency \cite{WernerTimestepConstraints}. The stability requirements of the Leapfrog method \cite{TimeStabilityConditions} requires mesh cell sizes smaller than the electron Debye length. As a result, the numerical requirements of fully kinetic PIC simulations are dominated by the requirements involved in simulating electron motion. The use of hybrid fluid-kinetic simulations, in which electrons are simulated as a Boltzmann electron fluid, and ions are simulated directly, lowers the numerical requirements of the simulation, at the small cost of solving for the electron fluid distribution.\\
In pdFoam, the electron distribution was assumed to be described by an isothermal, unmagnetised, inertia-less ($\frac{m_e}{m_i}\rightarrow 0$) electron fluid  \cite{Whipple81},\cite{Stone1981SpaceAerodynamics}. Under these assumptions, the magnetohydrodynamic equations of continuity, momentum, and energy reduce to Eqn.\ref{Electron Fluid} \cite{BoltzmanEFLuid}:
\begin{equation}
    n_e=n_{e,\infty} exp \left[ \frac{q_e \phi(\bf{x})-\phi_{\infty}}{k_BT_e}\right]
    \label{Electron Fluid}
\end{equation}
 where $k_B$ is the Boltzmann constant, $T_e $ is the electron temperature, $n_{e,\infty}$ is the freestream electron number density, and $\phi_{\infty}$ is the freestream potential.\\

\section{pdFoam simulation layouts}
\begin{figure}[!htb]
         \centering
             \includegraphics[width=0.95\textwidth]{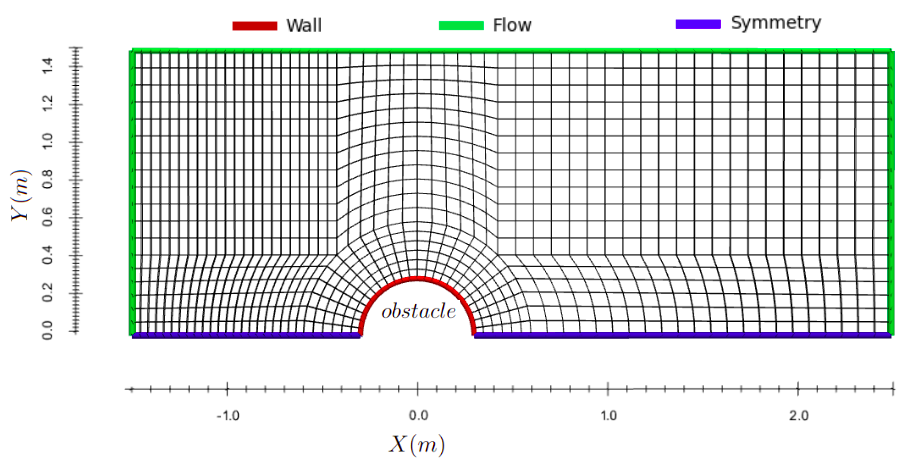}
         \caption{Grid structure employed in simulating flow over small satellites, with $obstacle$ patch marked, and boundary conditions on faces indicated by coloured lines. Grid is only one cell long in z direction, so has been displayed in 2d, neglecting the $front$ and $back$ faces (which also employ symmetry boundary conditions)}
         \label{BoundaryConditions}
         \end{figure}

To simulate a satellite in mesothermal flow, the approximation of small satellite geometry as a cylinder was used. The principle advantages of a cylinder are its relative simplicity, and the ability to directly compare with orbital motion limited theory (Outlined in Chapter \ref{C:wakeStructures}) 
The simulation was performed on the grid shown in Fig.\ref{BoundaryConditions}, representing the top half of the region being simulated (symmetry boundary conditions were employed to lower computational load), with the $obstacle$ patch marked, and the respective boundary conditions (BC) as described here.

\begin{figure}[!ht]
         \centering
             \includegraphics[width=0.9\textwidth]{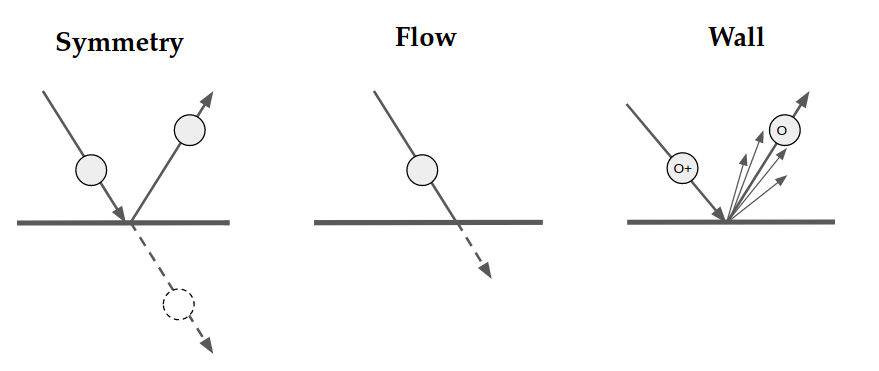}
        \caption{Example particle trajectories as particle passes across the three boundary conditions, showing specular reflection, deletion, and diffuse reflection/neutralisation, respectively}
         \label{BoundaryConditionsExplanation}
         \end{figure}

As the simulation is symmetrical along the $xz$ axis (through $(0,0,0)$), and along the $yz$ axis, the bottom patch was rendered with symmetry boundary conditions, as were the front and back patches. the simulation was performed with only one cell in the z direction, but particles velocities and positions retained their respective z component, allowing them to pass into and out of the symmetry regions on the $front$ and $back$ faces. \\
When a particles motion passes through a symmetry patch, such as the $bottom$ patch, it is returned to the simulation with the component of its velocity in the direction of the outward face reversed (as in Fig.\ref{BoundaryConditionsExplanation}) as if a particle with symmetry along that face was passing in from the other side. The patch marked $flow$ is treated with the flow boundary condition, and $obstacle$ face with corresponding obstacle condition. At each timestep, particles that pass out of faces with flow BC are removed, while a number of new macroparticles  given by 
\begin{equation}
    N_{k,new}=A_{\perp} \cdot v_{flow}dt \left(\frac{\rho_{N,k}}{N_{Equiv}}\right)
    \label{NewMacroCalculation}
\end{equation}
is introduced, where $A_{\perp}$ is the cross sectional area of the face to the flow direction, $\rho_{N,k}$ and $v_{flow}$ are the number density and magnitude of flow velocity for species $k$, and all other parameters are as defined in Appendix \ref{C:NotationGuide}.
When particles interact with faces using the obstacle BC, they are reflected diffusely, and charged particles are converted to their respective neutral particle. the momentum change in particles reflecting from the $obstacle$ patch is recorded and summed for each face, to be used in determining direct drag force on the object.
The $flow$ face is held at a constant potential $\phi$ determined by initial conditions (0V in this case), while symmetry faces boundary conditions have potential gradient set to zero, and the $obstacle$ face has a fixed potential (denoted as $\phi_0$, usually -50V in this thesis)
The mesh density was determined alongside other key initial parameters with a numerical convergence study laid out in Chapter \ref{C:NumConv}.

\section{Control surface approach and the Maxwell stress tensor}

In order to identify  steady state solution to these simulations, it is necessary to collect a large range of data as a function of time. Large datasets cause it to be grossly inneficient for this data to be recorded for anything other than the final, steady state conditions. To fulfill the requirements of monitoring the simulation at each timestep, it is therefore necessary to implement secondary methods, to record representative values  for each timestep, separate to the writeFiles stage (In which the final state of the simulation is recorded for all fields). In the typical simulation, the total simulated time is 0.001s, while the algorithm iterates in timesteps $\delta t$ of $10^{-7}$s, so it impractical to record much more than a handful of representative values. Being the primary interest of this investigation, those values were chosen to be the total direct force on the satellite, and the charged (indirect) force.
The direct (neutral) force on the satellite can be obtained as a sum of force moments from diffusely reflected particles, a functionality which is built in to the pdFoam solver already, providing net direct force values for each timestep.Taking a discrete approach to charged force calculation would be impractical

To determine the charged drag on the cylinder, we begin by considering the general momentum balance of a mesothermal plasma-body interaction laid out by \cite{ControlSurfaceAllen}. This will be demonstrated with the inclusion of a magnetic field, although this will later be neglected under electrostatic assumptions for these simulations. 
In a mesothermal flow, the thermal ion and electron pressures  are assumed to be negligible compared to the streaming energy, letting us write the ion and electron momentum equations as 
\begin{equation}
    n_im_i\left[ \frac{\partial \mathbf{v}_i}{\partial t} +(\mathbf{v}\cdot\nabla)\mathbf{v}_i \right]=q_in_i(\mathbf{E}+\mathbf{v}_i\times\mathbf{B})
    \label{IonMomentum}
\end{equation}
\begin{equation}
    n_em_e\left[ \frac{\partial \mathbf{v}_e}{\partial t} \right]=q_en_e(\mathbf{E}+\mathbf{v}_e\times\mathbf{B})
    \label{Electron momentum}
\end{equation}
 Where $\mathbf{E}$ and $\mathbf{B}$ are the electric and magnetic field vectors, $n_i$ and $n_e$ the ion and electron number density, and $\mathbf{v}$ the flow velocity \cite{GriffithsElectrodynamics}. Taking the sum of these two equations, and introducing charge density as $\rho_c=(\sum_k^K Z_k q_e n_k)-q_en_e$ and current density $\mathbf{J}=\rho_c\mathbf{v}$ gives
\begin{equation}
    (m_in_i+m_en_e)\frac{\partial \mathbf{v}}{\partial t}+m_in_i(\mathbf{v}\cdot\nabla)\mathbf{v} =\rho_c \mathbf{E}
+\mathbf{J}\times\mathbf{B}    \label{SumOfMomentums}
\end{equation}

Next we introduce the divergence form of Gauss' law and the Maxwell-Ampere equation to express $\rho_c$ and $\mathbf{J}$ in terms of $\mathbf{E}, \mathbf{B}$
\begin{equation}
    \rho_c=\epsilon_0 \nabla\cdot\mathbf{E}
    \label{gaussDiverge}
\end{equation}
\begin{equation}
    \mathbf{J}=\frac{1}{\mu_0}(\nabla\times\mathbf{B})-\epsilon_0\frac{\partial\mathbf{E}}{\partial t}
    \label{MaxWellAmpereDivergence}
\end{equation}
Equations \ref{gaussDiverge} and \ref{MaxWellAmpereDivergence} can be substituted into Eqn.\ref{SumOfMomentums} to give 
\begin{equation}
    (m_in_i+m_en_e)\frac{\partial \mathbf{v}}{\partial t}+m_in_i(\mathbf{v}\cdot\nabla)\mathbf{v}=\epsilon_0(\nabla\cdot\mathbf{E})\mathbf{E}+\frac{1}{\mu_0}(\nabla\times\mathbf{B})\times\mathbf{B}-\epsilon_0\frac{\partial \mathbf{E}}{\partial t} \times \mathbf{B}
    \label{substitutionofGauss}
\end{equation}
using the product rule and Faradays law, the time derivative of the electric field can be written in terms of $\partial/\partial t (\mathbf{E}\times \mathbf{B})$\cite{CaponBible}
\begin{equation}
    \ddt{ }(\B{E}\times\B{B})=\ddt{\B{E}}\times\B{B}+\B{E}\times\ddt{\B{B}}=\ddt{\B{E}}\times\B{B}-\B{E}\times(\nabla\cdot\B{E})
    \label{Identity103}
\end{equation}
substituting Eqn.\ref{Identity103} into Eqn.\ref{substitutionofGauss} gives

\begin{equation}
\begin{split}
    (m_in_i+m_en_e)\ddt{\B{v}}+m_in_i(\B{v}\cdot\nabla)\B{v}&=\epsilon_0[(\nabla\cdot\B{E})\B{E}-\B{E}\times(\nabla\times\B{E})]
    \\
    +\frac{1}{\mu_0}[(\nabla\cdot\B{B})\B{B}-&\B{B}\times(\nabla\times\B{B})]-\epsilon_0\ddt{ }(\B{E} \times \B{B})
\end{split}
    \label{BadEquation}
\end{equation}
The curls in Eqn.\ref{BadEquation} can be eliminated by using 
\begin{equation}
    \B{A}\times(\nabla\times\B{A})=\frac{1}{2}\nabla(\B{A}\cdot\B{A})-(\B{A}\cdot\nabla)\B{A}
    \label{VCalcIdentity}
\end{equation}
Eqn.\ref{BadEquation} then becomes
\begin{equation}
\begin{split}
    (m_in_i+m_en_e)\ddt{\B{v}}+m_in_i(\B{v}\cdot\nabla)\B{v}&=\epsilon_0[(\nabla\cdot\B{E})\B{E}+(\B{E}\cdot \nabla)\B{E})]
    \\
    +\frac{1}{\mu_0}[(\nabla\cdot\B{B})\B{B}+&(\B{B}\cdot\nabla)\B{B}]-\epsilon_0\ddt{ }(\B{E} \times \B{B})
\end{split}
    \label{BadEquationUncurly}
\end{equation}
This can be simplified considerably with the use of the Maxwell stress tensor $\B{\overline{T}}$
\begin{equation}
 \B{\overline{T}_{i,j}}=\en\left(E_iE_j-\frac{1}{2}\delta_{i,j}E^2\right)+\frac{1}{\mu_0}\left(B_iB_j-\frac{1}{2}\delta_{i,j}B^2\right)
    \label{StressTensor}
\end{equation}
where $\delta_{i,j}$ is a kronecker delta. taking the divergence of $\B{\overline{T}}$ and introducing the Poynting vector $\B{S} =\mu_0^{-1}(\B{E}\times\B{B})$, Eqn.\ref{BadEquationUncurly} can now be written as 
\begin{equation}
m_in_i(\B{v}\cdot\nabla)\B{v} -\nabla\B{\overline{T}}=-(m_in_i+m_en_e)\ddt{\B{v}}-\en\mu_0\ddt{\B{S}}
\label{ShortenedBadEquation}
\end{equation}
For a steady state solution, we allow the time derivatives terms on the RHS to tend to zero. To obtain a control surface formulation of the resulting equation, we then consider surface $S$ containing volume $V$, defined by outward normal vector $\B{\hat{n}}$,, such that equation \ref{ShortenedBadEquation} becomes
\begin{equation}
\int_S n_im_i(\B{v}\cdot\B{\hat{n}})\B{v}dS - \int_S \B{\overline{T}}\cdot\B{\hat{n}}dS=0
\label{ControlSurfaceFirst}
\end{equation}
Where the two integrals can be taken to represent ion momentum, and the electromagnetic stress (from left to right) respectively. This equation implies that the variation in mechanical momentum between flows into and out of the volume are caused by net electromagnetic (and mechanical) interactions within the volume (conservation of momentum in an electromagnetic field must account for the momentum stored by the fields).
The influence of the plasma on a body can then be found by defining a second surface $S_2$ to represent the charged body, such that 
\ref{ShortenedBadEquation} becomes
\begin{equation}
\int_S (n_im_i(\B{v}\cdot\B{\hat{n}})\B{v} -  \B{\overline{T}}\cdot\B{\hat{n}})dS+\int_{S_2}( n_im_i(\B{v}\cdot\B{\hat{n}})\B{v} -  \B{\overline{T}}\cdot\B{\hat{n}})dS_2=0
\label{ControlSurfaceSecond}
\end{equation}
Where the second integral is the force exerted by a plasma on the body defined by $S_2$. This gives the force on a surface S as
\begin{equation}
\B{F_c}=-\int_Sn_im_i(\B{v}\cdot\B{\hat{n}})\B{v}dS+\int_S\B{\overline{T}}\cdot\B{\hat{n}}dS
\label{ControlSurfaceLast}
\end{equation}
The first term in this is analogous to the discrete summation method employed in calculating direct force, replacing a summation of individual momentum exchanges with an integral over the average mechanical momentum exchange. The second term provides a convenient way to calculate indirect force for any face on the body surface, requiring only the electric field and surface normal at that point.

\section{Indirect force calculation}\label{section:indirectForceCalculation}

\begin{figure}[h!]
         \centering
     
         \begin{minipage}[c]{0.8\textwidth}
             \centering
             \includegraphics[width=0.61\textwidth]{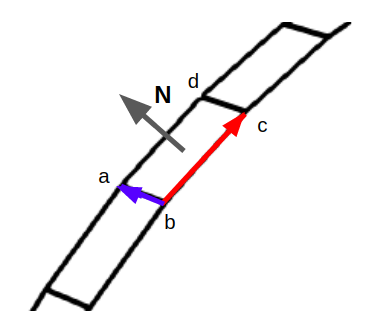}
              \centering
         \end{minipage}
        \captionsetup{width=0.51\linewidth}
         \caption{Calculation of the normal area vector from the vertex points of a face element}
         \label{CrossProductArea}
     \end{figure}

In order to implement the indirect force calculations, the integral in Eqn.\ref{ControlSurfaceLast} is converted to a summation, for the program to perform at each step of the simulation.
\begin{equation}
    \textbf{f}_c=\sum_{i}^S \textbf{T}_i \cdot \B{n}_i
    \label{IndirectForceSum}
\end{equation}
Where the index i represents a face in the surface S, and $\B{n}_i$ is the normal vector for this face. In order to calculate each entry in this integral, the electric field is extracted  (in our case we used the electric field calculated in the potential assignment step of the algorithm in Fig.\ref{FlowChart}), over the surface of interest (the $obstacle$ patch in this case). The value for $\hat{n}$ can be obtained with a loop over the surface of interest, taking each face element, and for a face defined by points (a, b, c, d), taking the normal area vector as the cross product $N=\vec{ba}\times\vec{bc}$ shown in  Fig.\ref{CrossProductArea}. This can be done with a loop over the face set for the surface of interest and the 'SArea' method in openFoam.\newpage
Two approaches were considered in the method of calculating the product $\textbf{T}_i \cdot \hat{n}_i$, with the first being the piecewise definition, in which the $x, y, z$ elements of $\hat{n}$ and $\textbf{E}$ are extracted, and the Indirect force for each face is calculated as the vector with entries $(F_{E,x}, F_{E,y}, F_{E,z})$ (with subscript $C$ used instead of $i$ to clearly distinguish from index i in Eqn.\ref{IndirectForceSum}):

\begin{equation}
\begin{aligned}
F_{C,x}=(E_x^2-\frac{\textbf{E}^2}{2})n_x+E_xE_yn_y+E_xE_zn_z \\
F_{C,y}=E_xE_yn_x+(E_y^2-\frac{\textbf{E}^2}{2})n_y+E_yE_zn_z \\
F_{C,z}=E_xE_zn_z+E_yE_zn_z+(E_z^2-\frac{\textbf{E}^2}{2})n_z
\end{aligned}
    \label{IndirectForcepiecewise}
\end{equation}     

\begin{figure}[h!]
         \centering
         \begin{minipage}[c]{0.95\textwidth}
             \centering
             \includegraphics[width=\textwidth]{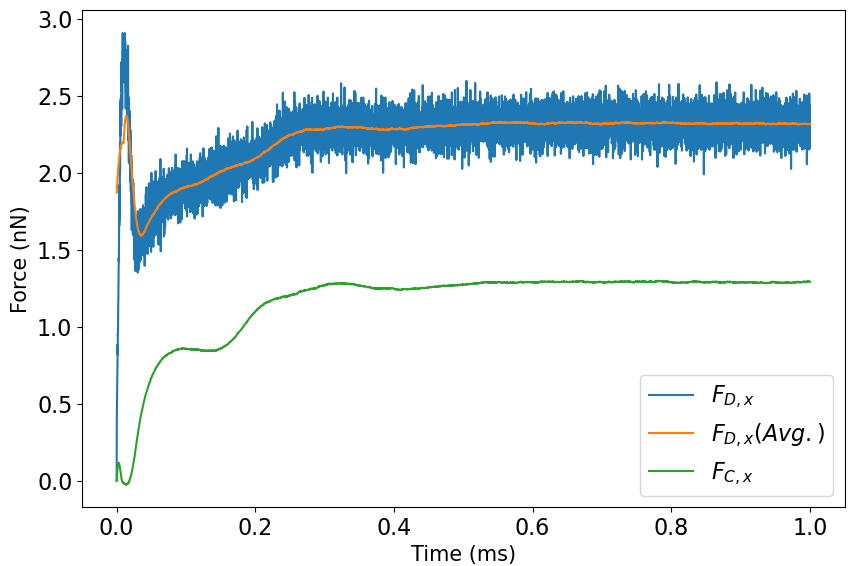}
              \centering
         \end{minipage}
        \captionsetup{width=0.91\linewidth}
         \caption{Total direct force $F_{D}$ and charged force $F_C$  on the cylinder (in direction of flow) as a function of time. $F_D$ is calculated from collisional momentum exchanges between particles and the cylinder, while $F_C$ is calculated using the method laid out above, in section \ref{section:indirectForceCalculation}. A running average has been overlaid on the direct force for clarity ($F_{D}$(Avg.))}
         \label{ForcesSteadyState}
         \end{figure}

This method suffers from inefficiencies associated with converting the vector fields into 3 scalar fields each, then converting them back after performing the additions. The second approach to this calculation solves these efficiency issues by keeping $\textbf{E}$ and $\hat{n}$ as vector fields, and performing the calculation through openFoam vector manipulation methods instead. This gives the expression:
\begin{equation}
    \textbf{F}_C=(\textbf{E}\textbf{E}^T-\frac{1}{2}{\bf{\mathbf{1}}}|\textbf{E}|^2)\cdot \B{n}
    \label{IndirectForceVectorMagic}
\end{equation}
This vector based solution structure allows for both more compact code, and more efficient and scalable solutions, making it the ideal choice to meet the constraints of the indirect force monitoring code.\\
This method was implemented as an optional method in pdFoam, allowing it to be removed in cases where the additional data was not worth the increased computational cost (Although the computational cost of this method is $ <1\%$ of the cost of the charge assignment method alone in the meshes used here, so this is not significant enough to ever justify disabling this method). A sample of output data is shown in Fig.\ref{ForcesSteadyState}, in which the average of charged and direct force in the x direction for each timestep is plotted over the raw values. Notably, the charged force has considerably less noise than that of the direct force, as the direct force is determined by particles impacting the cylinder in one time interval, while the charged force is determined by the local potential, which shows much less time variation than a measurement of particle density at a point. The sharp rise at the start of the plot is a feature of the formation of the sheath, with the quasi-neutral plasma (which provides no charged force, and allows little room for ions to accelerate, leaving a reduced direct force) in the initial conditions separating to form a plasma sheath (steady state condition).

\chapter{Numerical Convergence}\label{C:NumConv}
\section{Convergence parameters and bounding values}

A numerical model is convergent if a sequence of model solutions with increasingly refined conditions approaches a fixed value \cite{ConvergenceConditionsThompson}. Likewise, a numerical model is consistent if and only if the sequence converges to the solution of the physical equations governing this particular phenomena. In the case of PIC simulations, the key approximations made are the discretization of the timestep, the mesh, and the macroparticle approximation, splitting continous space and time into discrete timesteps and cells for the algorithm to iterate through, and replacing a large number of discrete particles with a smaller number of macroparticles (all of which are outlined in Chapter \ref{C:pdFoam}.)\\
This means that if this model is both convergent and consistent, the solution given by our simulations should approach the real value of the systems modelled as the cell size and timestep are lowered, and the number of macroparticles used is raised.
As a finer mesh and smaller timestep is used, the computational cost increases accordingly, and it is the goal of the Numerical convergence study undertaken here to determine a set of parameters for an optimal balance of computational cost to simulation fidelity to be used in further simulations. \par
\begin{figure}[h!]
         \centering
     
         \begin{minipage}[c]{0.95\textwidth}
             \centering
             \includegraphics[width=0.95\textwidth]{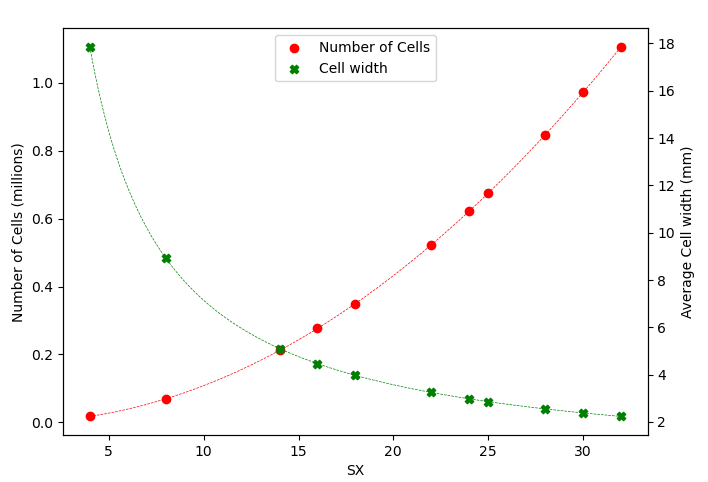}
              \centering
         \end{minipage}
        \captionsetup{width=0.7\linewidth}
         \caption{Total cells in the mesh, and average cell width, vs SX parameter. The runtime of the field solver scales with the number of mesh elements squared \cite{FEMScaling}, while the Courant number scales with $1/\Delta x$, so a compromise must be reached between these two conditions}
         \label{SX_CellParameters}
     \end{figure}

For this study, the mesh density was scaled by a parameter SX, where an SX of 1 corresponds to the mesh shown in Fig.\ref{BoundaryConditions}, and the number of cells in each of the $x$ and $y$ directions scales roughly linearly with an increasing SX, giving a relationship between total number of cells and SX parameter shown in Fig.\ref{SX_CellParameters}. The time interval for each step of the algorithm was denoted $\Delta t$, while the ratio of real particles to simulated particles was denoted $N_{Equiv}$. 
The main bounds imposed on these parameters were those imposed by computational cost at the fine end, and the Courant Condition at the coarse end. The Courant condition is a necessary condition for the convergence of the solutions of the PDEs laid out in Chapter \ref{C:DimensionlessParameters}. It states that for a given flow velocity $u$, spatial dimension $\Delta x$ and time interval $\Delta t$,
\begin{equation}
    C=\frac{u \Delta t}{\Delta x} \leq C_{max}
    \label{Courant condition}
\end{equation}
Where $C_{max}=1$ is usually used for a time stepping solver. Physically, this means that any propagating wave with velocity $u$ will spend at least one time step in each consecutive cell on its path, allowing the behaviour of these signals to be resolved.

\section{Mesh and Particle density convergence}
\begin{figure}[h!]
         \centering
     
         \begin{minipage}[c]{0.9\textwidth}
             \centering
             \includegraphics[width=0.9\textwidth]{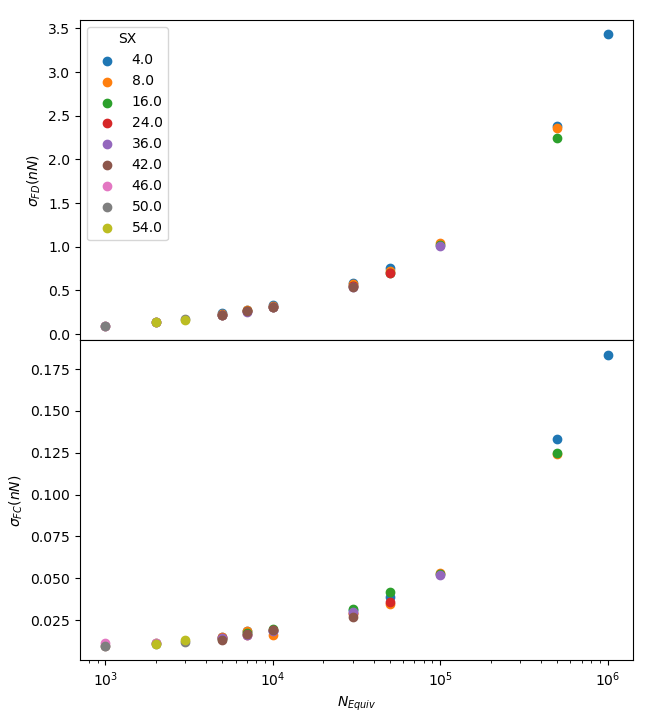}
              \centering
         \end{minipage}
        \captionsetup{width=0.9\linewidth}
         \caption{Standard deviation of Direct and charged forces as a function of the $N_{Equiv}$ parameter, showing strong convergence as $N_{Equiv}\rightarrow 1$, and very little variation between different SX values. }
         \label{STDFD_byNEQ_Subplots}
     \end{figure}

A range of cases were prepared, with conditions laid out in table below, varying both the SX and $N_{Equiv}$ parameters, in the range 4 - 56 and $10^3$ - $10^6$ respectively.
\begin{center}
\begin{tabular}{||c | c c ||} 
 \hline
 Constant & Value & Unit\\ [0.5ex] 
 \hline\hline
 $v_{\infty}$ & $7500$ & $ms^{-1}$\\ 
 \hline
 $\rho_{H+}$ & $2\times10^{10}$ & $m^{-3}$\\
 \hline
 $\rho_{O+}$ & $2\times10^{10}$ & $m^{-3}$\\
 \hline
 $\phi_B$ & $-50$ & $V$ \\
 \hline
 $r_0$ & 0.3 & $m$  \\ [1ex] 
 \hline
\end{tabular}
\end{center}
These cases were run to a steady state, and the charged and direct force in the x direction were measured using the methods in section \ref{C:pdFoam}. In the case of mesh parameters and Macroparticle convergence, the average force values varied significantly enough between similar cases that the standard deviation ($\sigma$) of force values for the last 50$\%$ of the simulation was used as a surrogate value to quantify convergence. $\sigma$ was defined as in Eqn. \ref{StandardDevEquation}, where $f_t$ is the integrated force at timestep $t$,  N is the total number of timesteps sampled, and $\mu$ is the mean . As the number of timesteps taken after each case reaches steady state is constant between the cases, the $\sigma$ values for that case become an effective measure of quantifying variation within the force values.
\begin{equation}
    \sigma_f=\sqrt{\frac{\sum_t (f_t-\mu)^2}{N}}
    \label{StandardDevEquation}
\end{equation}
As the set of parameters grows arbitrarily fine, the value of $\sigma$ approaches a converged value $\sigma_0$. The criteria chosen to qualify a set of parameters as adequately converged was the value of $\sigma$ corresponding to those parameters being within 10$\%$ of $\sigma_0$. Fig.\ref{STDFD_byNEQ_Subplots} shows variation of $\sigma_0$ with SX, for a number of different $N_{Equiv}$ values.
\begin{figure}[h!]
         \centering
     
         \begin{minipage}[c]{0.95\textwidth}
             \centering
             \includegraphics[width=\textwidth]{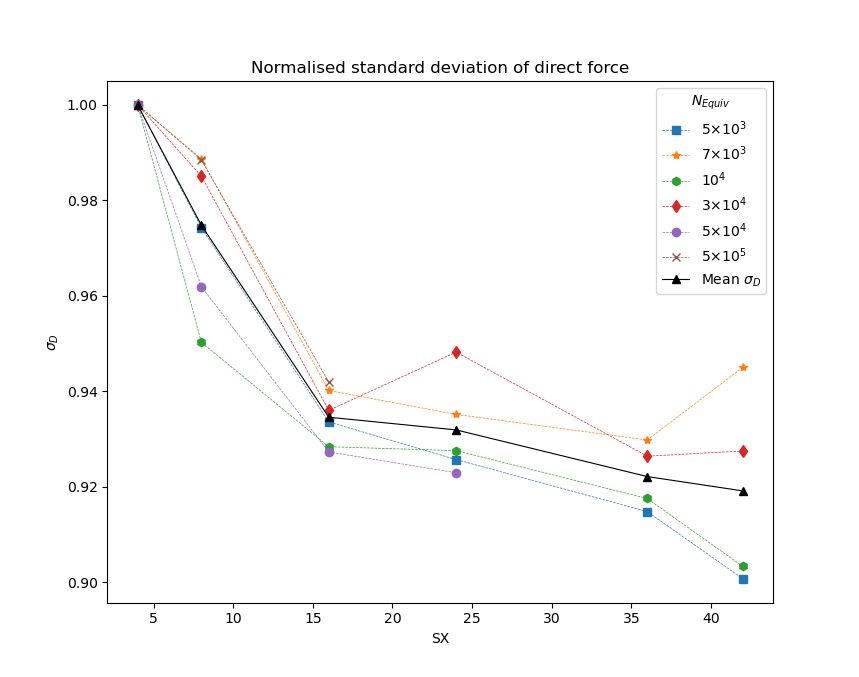}
              \centering
         \end{minipage}
        \captionsetup{width=0.7\linewidth}
         \caption{Standard deviation of direct force as a function of the SX parameter for a representative selection of $N_{Equiv}$ values. }
         \label{STDFDX_bySX}
     \end{figure}
     \newpage
Analysis of the $\sigma_F$ values for the charged and direct force showed a strong convergence with changing $N_{Equiv}$,  (Fig.\ref{STDFD_byNEQ_Subplots}) (given by use the coarsest set of parameters), while an increasing SX value only accounted for a convergence of 10$\%$ or less of the maximum value (Fig.\ref{STDFDX_bySX}), with more than half of this variation occuring in the range SX$<16$. As the normalised $\sigma$ values taken above this mesh parameter seem to not be significantly affected by changing SX, the mesh was considered to be converged at an SX of 30 (and a value of SX=40 was chosen to be used in final simulations, allowing better resolution of flow features in the fore-wake). \par
The standard deviation values showed a strong convergence as $N_{Equiv}$ decreased, in both the direct and indirect forces (Fig.\ref{STDFD_byNEQ_Subplots}). To identify the point at which $\sigma$ had converged to within the $10\%$ margin selected, the curves for $\sigma_D$ and $\sigma_C$ were normalised (setting the maximum value of each curve to 1), and fit to an empirical function for ease of interpolation. The function chosen was given by $\sigma=A/x^n+\sigma_0$, for arbitrary scaling parameter A, final converged standard deviation $\sigma_0$, and variable of interest $x$. \newpage

The data and fit were presented in terms of total number of macroparticles, rather than $N_{Equiv}$, as the simulations of different densities aim to keep the same number of macroparticles per cell, by adjusting the $N_{Equiv}$ parameter accordingly.

\begin{figure}[h!]
         \centering
     
         \begin{minipage}[c]{0.9\textwidth}
             \centering
             \includegraphics[width=0.9\textwidth]{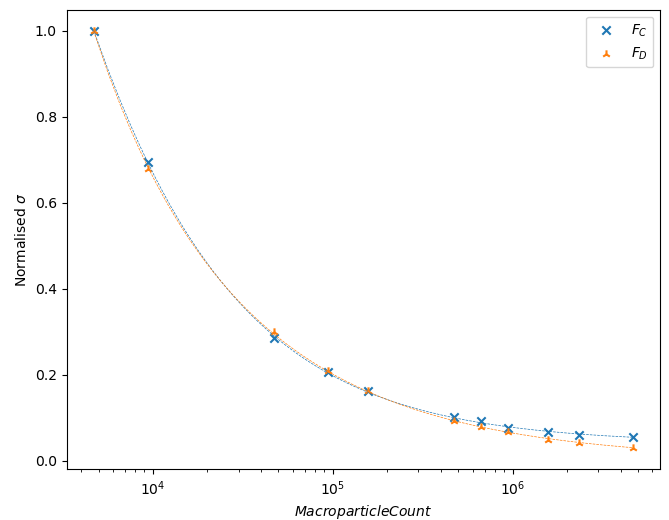}
              \centering
         \end{minipage}
        \captionsetup{width=0.7\linewidth}
         \caption{Normalised $\sigma$ for both  direct and indirect force, presented as a function of the total macroparticles in the steady state simulation.}
         \label{STDX_byMacroparticles_fit}
     \end{figure}
In Fig.\ref{STDX_byMacroparticles_fit}, the final value $\sigma_0$ towards which $\sigma$ converges was found to be $10^{-17}N\approx 0$ in the direct case, and  $0.004N$ in the indirect case, implying the standard deviation approaches zero as the number of macroparticles rises to infinity  (Realistically, this is not the case, as the number of macroparticles used is bounded by the upper limit where each macroparticle represents one real particle). This fit allows the identification of the $10\%$ margin as being  $\approx 4.5\ex{5}$ macroparticles.

\section{Optimal time step convergence}
\begin{figure}[h!]
         \centering
     
         \begin{minipage}[c]{0.9\textwidth}
             \centering
             \includegraphics[width=0.9\textwidth]{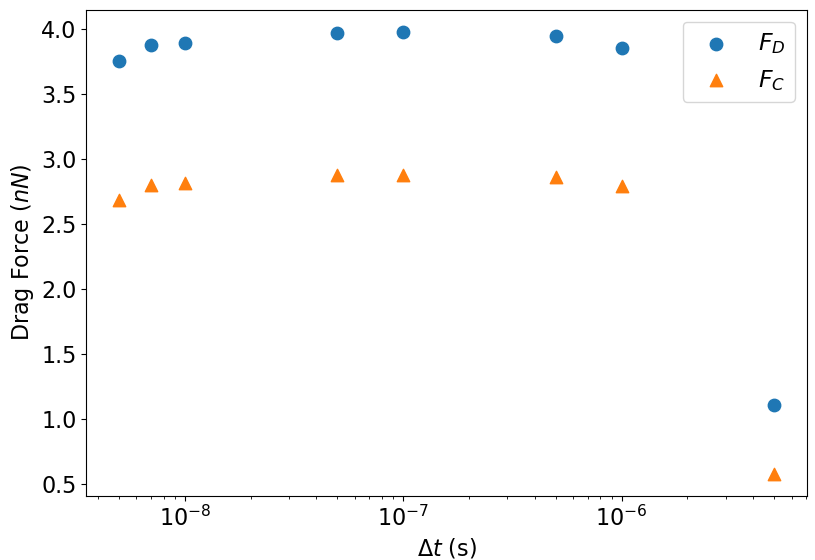}
              \centering
         \end{minipage}
        \captionsetup{width=0.7\linewidth}
         \caption{Direct and Charged drag force averages for cases with varying time intervals, showing a close grouping for values of $\Delta t<10^{-6}$, and a strong divergence above this value.}
         \label{FbyDTConvergence}
     \end{figure}

A selection of cases was prepared with these new values for SX and $N_{Equiv}$, with the time step for one iteration of the algorithm varying from $5\times 10^{-6}$s to $5\ex{-9}$s between cases. Each case was run until a time of 0.001s as before, upon which convergence was complete 
When measuring convergence of the time step $\Delta t$, the standard deviation becomes an inneffective tool, as the number of timesteps being sampled increases as $\Delta t$ decreases, resulting in the standard deviation growing with a finer timestep, rather than converging.
As a result, the direct and charged force averages must be used directly, resulting in Fig.\ref{FbyDTConvergence}. It is immediately obvious that the dominant source of variation for timesteps of $10^{-6}$s or smaller is random variation within the simulation, while the timestep of $5\ex{-6}s$ shows significant variation from these, in both charged and direct force values. This is to be expected, as the average cell size is around 1mm, so the courant number for the main flow of this simulation would be $C\approx10$, well above the convergence criteria $C\leq1$. While the force values appeared to converge for $\Delta t =10^{-6}s$, the time step used in further simulations was chosen to be $10^{-7}$s, allowing the courant condition to still be satisfied for particle speeds of up to 70$\%$ greater magnitude than the bulk flow, such as those in the close fore-wake region. 

\chapter{Dimensional analysis}\label{C:DimensionlessParameters}

A common technique in modelling complex aerodynamic systems is the reduction of highly dimensional, complex systems into representative dimensionless parameters \cite{HallPlasmaVehicleInteraction}\cite{BeiserScalingLaws}\cite{knechtel1964experimental}\cite{PlasmaExpansionIntoVacuumSamir}. Previous work by \cite{Lacina_1971} applied the Buckingham Pi theorem to microscopic motion of ions in a plasma, as governed by the Vlasov-Maxwell equations. Other work (\cite{BeiserScalingLaws}) used the Buckingham Pi theorem to find dimensionless similarity parameters in a dielectric, conducting, viscous medium. These systems are suited to explaining the response of a plasma to a disturbance, but fall short in describing the aerodynamic interactions of a plasma and a solid body. An alternative set of parameters for ionospheric aerodynamic analysis is given by Capon et al \cite{CCScaling}, summarised here.
\section{Dimensionless Parameters}
This derivation functions on the assumptions of electrostatic interactions of a collisionless, unmagnetised multi-species plasma with multiply charged ions with a body at a fixed potential with respect to a quasi-neutral freestream plasma. Electron distribution is modelled as a Boltzmann electron fluid \cite{Whipple81}, an isothermal, inertia-less electron fluid.
Quantities referencing the satellite body are denoted with the subscript $0$, while quantities referring to the main flow (without perturbations) are denoted with $\infty$.
\par
As in section \ref{C:pdFoam}, Equations \ref{Boltzmann equation} and \ref{MaxWellAmpereDivergence} describe the evolution of a system in terms phase space distribution function $f$, potential $\phi$, and space-charge density $\rho$ given by 
\begin{equation}
    \rho=\left(\sum_k^K q_kn_k\right)-q_en_e \quad n_k=\int f_kd\mathbf{c}
    \label{SecondChargeDensity}
\end{equation}
where the electron density $n_e$ in the Boltzmann electron fluid model is described by
\begin{equation}
    n_e=n_{e,\infty} exp\left[\frac{q_e\phi(x)}{k_BT_e}\right]
    \label{BoltzmannElectronFluid}
\end{equation}
Where $n_{e,\infty}$ is given as the sum of the respective ionic charges in a quasi-neutral freestream (as net charge density in freestream=0).
\begin{equation}
    n_{e,\infty}=\frac{1}{q_e} \sum^K_k q_kn_{k,\infty}
    \label{freestreamElectronDensity}
\end{equation}
substituting the expressions for charge densities from Eqns \ref{SecondChargeDensity}-\ref{freestreamElectronDensity} into equation \ref{MaxWellAmpereDivergence} gives a pair of expressions describing distribution of a $K$ species plasma denoted by subscript $k$
\begin{equation}
    \ddt{f_k}+\B{c}_k\cdot\nabla_xf_k-\frac{q_k}{m_k}\nabla_x\phi\cdot\nabla f_k=0
    \label{DimensionedEqn1}
\end{equation}
\begin{equation}
    \nabla^2\phi=-\frac{q_e}{\epsilon_0}\sum^K_k\frac{q_k}{q_e}\left(n_k-n_{k,\infty}exp\left[\frac{q_e\phi}{k_BT_e}\right]\right)
    \label{DimensionedEqn2}
\end{equation}
To express Eqns \ref{DimensionedEqn1}-\ref{DimensionedEqn2} in dimensionless form, for each parameter $\psi$ with units $U_n$, a substitution is made such that $\Psi=\frac{\psi}{\psi_0}$, where $\psi_0$ is an characteristic value for the corresponding parameter, also using units $U_n$, and $\Psi$ is a dimensionless variable corresponding to that unit. As an example, the velocity $\B{c}_k$ becomes $\B{c}_k=v_{k,0}\B{u}_k$, where $v_{k,0}$ is a representative velocity (freestream velocity in this case), and $\B{u}_k$ is a dimensionless representation of velocity. This allows measurement with respect to the characteristic values, so $\B{u}_k$ is 1 in the freestream, and  greater than 1 in regions in which it has experienced an acceleration, respectively. The full list of substitutions is given here:

\begin{align*}
\B{x}=&r_0\B{X} & t=&\tau/\omega_0 \\ \B{c}_k=&v_{k,0}\B{u}_k &
f_k&=n_{k,\infty}F_k, \\ \phi&=\phi_0\B{\Phi}, & q_k&=q_eZ_k\\
n_k=&n_{k,\infty}N_k, & \nabla_x=&\hat{\nabla}_X/r_0 \\   \nabla_c=&\hat{\nabla}_u/v_{k,0} 
\end{align*}
For  $r_0$, $\omega_0$, $v_0$, $\phi_0$ as characteristic length, frequency, velocity, and potential respectively. $\B{X},\tau,\B{u},F,\Phi,Z,$ and $N_k$ are dimensionless parameters, and $\nabla_x$, $\nabla_u$ are the dimensionless gradient operators in physical and velocity space, respectively. Substituting these transformations into Eqn.\ref{DimensionedEqn1}, Eqn\ref{DimensionedEqn2} gives
\begin{equation}
    \left(\frac{\omega_0 r_0}{v_{k,0}}\right)\ddt{F_k}+\B{u}_k\cdot\hat{\nabla}_XF_k-Z_k\left(\frac{q_e\phi_0}{m_kv^2_{k,0}}\right)\hat{\nabla}_X\Phi\cdot\hat{\nabla}_uF_k=0
    \label{DimensionlessSubstitution1}
\end{equation}
\begin{equation}
    \hat{\nabla}^2_X\Phi=\sum^K_k-Z_k\frac{r^2_0q_en_{k,\infty}}{\epsilon_0\phi_0}\left(N_k-\exp{\left[\left(\frac{q_e\phi_0}{k_BT_e}\right)\Phi\right]}\right)
    \label{DimensionlessSubstitution2}
\end{equation}
The ion drift ratio is introduced as $S_k=v_{k,0}/v_{k,t}$, where $v_{k,0}$ is the ion drift velocity (the relative velocity between the charged body and the ion freestream), and $v_{k,t}$ is the ion thermal velocity, given by \cite{ThermalVel} 
\begin{equation}
    v_{k,t}=\sqrt{\frac{2k_BT_{k,\infty}}{m_k}}
    \label{Freestream velocity}
\end{equation}
allowing us to write Eqn.\ref{DimensionlessSubstitution1} as 
\begin{equation}
    \left(\frac{\omega_0 r_0}{v_{k,B}}\right)\ddt{F_k}+(1+S_k^{-1})\B{u}_k\cdot\hat{\nabla}_XF_k-Z_k\left(\frac{q_e\phi_0}{m_kv^2_{k,0}}\right)\frac{1}{(1+S_k^{-1})}\hat{\nabla}_X\Phi\cdot\hat{\nabla}_uF_k=0
    \label{DimensionlessSubstitution3}
\end{equation}
From inspection, there are 5 dimensionless parameters that govern the behaviour of Eqns \ref{DimensionlessSubstitution3}-\ref{DimensionlessSubstitution2},
\begin{align}
\alpha_k=&-Z_k\left(\frac{q_e\phi_0}{m_kv_{k,B}^2}\right),   &  \mu_e=&\left(\frac{q_e\phi_0}{k_BT_e}\right)\nonumber \\
\xi_k=&-Z_k\bfrac{r^2_0n_{k,\infty}q_e}{\epsilon_0\phi_0},     &  \Omega_k=&\bfrac{\omega_0r_0}{v_{k,B}}\label{DimParamList}\\
S_k=&v_{k,B}\sqrt{\frac{m_k}{2k_BT_{k,\infty}}} &&\nonumber
\end{align}
substituting the dimensionless parameters in Eqn.\ref{DimParamList} into Eqn.\ref{DimensionlessSubstitution2} and \ref{DimensionlessSubstitution3} gives 

\begin{equation}
\Omega_k \frac{\partial F_k}{\partial \tau}+(1+S_k^{-1})\B{u}_k\cdot\hat{\nabla}_XF_k+\frac{\alpha_k}{(1+S_k^{-1})}\hat{\nabla}_X\Phi\cdot\hat{\nabla}_uF_k=0
    \label{DimensionlessSubstitution4}
\end{equation}
\begin{equation}
\hat{\nabla}_X^2\Phi=\sum^K_k\xi_k(N_k-\exp[\mu_e\Phi])
    \label{DimensionlessSubstitution5}
\end{equation}
Reducing the 4+5K quantities ($\epsilon_0$, $q_e$, $T_e$, $k_0$, as well as separate $m_k$, $q_k$, $T_k$ and $n_k$ for each ion species k) to 1+4K dimensionless parameters ($\mu_e$, plus separate $\alpha_k$, $\xi_k$, $S_k$, and $\Omega_k$ for each ion species k.
The relative effect of each species on the potential $\Phi$ can be made more explicit by introducing the dimensionless parameters $\chi$ and $B_k$
\begin{equation}
\chi=\sqrt{\sum^K_k\xi_k}
    \label{ChiDef1}
\end{equation}
\begin{equation}
\beta_k=\frac{\xi_k}{\chi^2}
    \label{BetaDef1}
\end{equation}
Allowing Eqn.\ref{DimensionlessSubstitution5} to be expressed as 
\begin{equation}
\hat{\nabla}_X^2\Phi=\chi^2\left(\sum^K_k(\beta_kN_k)-\exp[\mu_e\Phi]\right)
    \label{DimensionlessSubstitution6}
\end{equation}
The physical meaning, and relative significance of these various parameters is laid out below

\section{Significance of scaling parameters}
Mathematically, the independent dimensionless parameters in \eqref{DimParamList} form a complete set of $\Pi$ groups to describe the K-species system of Vlasov-Maxwell equations \cite{CCScaling} (Eqn. \ref{DimensionedEqn1} and Eqn.\ref{DimensionedEqn2}). A physical intuition for each of these parameters is laid out here, with special attention paid to $\alpha_k$,$\beta_k$,$\chi$, the parameters of interest for this study.
\subsection{$S_k$: Ion Thermal ratio}
Given by the ratio of the the main freestream flow velocity to the average ion thermal speed, This parameter provides a measure of the thermal regime of the flow being simulated. In the case of mesothermal interactions, where $v_{t,i}\ll v_0\ll v_{t,e}$, $S_k\gg1$, and the evolution of distribution $F_k$ is uncoupled from thermal effects. In the opposite case, as the relative ion thermal velocity grows and $S_k\rightarrow 0$, the diffusion term in Eqn.\ref{DimensionlessSubstitution4}, $(1+S_k^{-1})\B{u}_k\cdot\DimD{X}F_k$ grows, and ion thermal energy becomes dominant in the evolution of $F_k$. An increased thermal velocity will also reduce the magnitude of the field effects term in Eqn.\ref{DimensionlessSubstitution4}, and the system will become decoupled from field effects and be entirely thermally dominated.
\\
In the case of LEO bodies, average orbital freestream velocities are on the order of 7-8km/s, while ion thermal velocities are on the order of 1km/s \cite{Banks2004ThermalVelOxy}, leading to $S_k$ of around 8 in most LEO cases. In this regime, thermal effects are unlikely to dominate any wake structures, but may result in more diffuse formations \cite{Whipple81}.

\subsection{$\mu_e$: electron energy coefficient}
$\mu_e$ is a common non-dimensional representation of potential \cite{GODDLaFramboiseCurrentCollection}\cite{knechtel1964experimental}. $\mu_e$ is the ratio between disturbance potential and electron thermal energy (analogous to $\alpha_k$ for electrons), and physically describes the penetration of the average electron into a potential barrier\cite{knechtel1964experimental}. In the Vlasov-Maxwell system, $\mu_e$ occurs in Eqn.\eqref{DimensionlessSubstitution6}, where it acts to scale the electron opposition to a potential disturbance. In this $\exp[\mu_e\Phi]$ term, we see that a high electron thermal velocity, and corresponding low $\mu_e$ will result in the local electron distribution being less tightly coupled to a disturbance in potential $\Phi$.

\subsection{$\Omega_k$: Ion Temporal parameter}
$\Omega_k$ is the ratio between a characteristic frequency for an ion disturbance, and the time for that disturbance to travel a characteristic length scale $r_0$. While not of particular interest in this study, this parameter acts to scale any change in temporal effects, so a system with smaller length scales, and therefore lower transit times, scales the frequency accordingly.
\subsection{$\beta_k$: Ion coupling parameter}
While not stricly one of the $\Pi$ groups, $\beta_k$ provides a more intuitive measure of multi-species plasma interactions. Using the explicit expression for $\xi_k$ in Eqn.\ref{DimParamList} in Eqn.\ref{BetaDef1}, we can re-write $\beta_k$ as

\begin{equation}
\beta_k=\frac{Z_kn_{k,\infty}}{\sum^K_kZ_kn_{k,\infty}}
    \label{BetaDef2}
\end{equation}
Which has the intuitive physical significance of the ratio of charge density from species k to the total charge density, or relative contribution to ion motion dominated effects of species k. An important property of $\beta_k$ for later derivations is the fact that $\beta_k$ can only range from 0 to 1, where a value approaching 0 implies that species k has a negligible effect on the total space-charge density, and a value of 1 implies that the response of the space charge density to a potential disturbance is dominated entirely by species k.

\subsection{$\alpha_k$: Ion deflection parameter}
$\alpha_k$ is the ratio of disturbance potential energy to kinetic energy of an ion, expressed as 
\begin{equation}
\alpha_k=\frac{1}{2}\frac{2Z_kq_e\phi_0}{m_kv^2_0}=-\frac{1}{2}\frac{P.E.}{K.E.}
    \label{alphaDef1}
\end{equation}
$\alpha_k$ acts on the external forces term in Eqn.\ref{DimensionlessSubstitution4} to scale the deflection of an ion by a potentential disturbance, relative to the ion's own momentum. $Z_k$ acts to control the direction of the deflection, with positive ions being deflected towards a negative disturbance and vice versa. In the case of the LEO interactions being studied here, the potential disturbances are rarely significantly positive ($\phi_0\leq 0$), and all ions are positively charged. As $|\alpha_k|\rightarrow0$, the ion kinetic energy becomes dominant over the disturbance potential, and ions experience negligible deflection, while in the opposite case, when $|\alpha_k|\rightarrow\infty$, the ion motion is dominated by the electric potential, and deflection is dominant over the original direction of motion.

\subsection{$\chi$: General shielding ratio}
The shielding ratio, or ratio between a characteristic length scale and the Debye shielding length, is a commonly used plasma parameter, given for a high temperature multi-species plasma (species denoted with k) as 
\begin{equation}
\chi=r_0\left(\frac{\sum_k^K Z_kn_{k,\infty}q_e^2}{\epsilon_0k_BT_e}\right)^{1/2}=\bfrac{r_0}{\lambda_{D}}
    \label{ChiDef2}
\end{equation}
In Eqn\eqref{DimensionlessSubstitution6}, the effect of $\chi$ can be seen as the ratio between potential disturbances in $\Phi$, and the corresponding response of the ion and electron distributions. As the body size grows relative to the shielding length, $\chi\rightarrow \infty$, and therefore the term $\DimD{X}^2\Phi\rightarrow\infty$, and a given disturbance $\phi_0$ will be strongly shielded, with the surrounding potential quickly approaching the average plasma potential.
As the body size shrinks with respect to the shielding length of the medium, and $\chi\rightarrow 0$, Eqn.\eqref{DimensionlessSubstitution6} becomes the Laplace equation $\nabla^2\Phi=0$, and electrical disturbances drop off proportional to $r^{-2}$. This particular formulation is only valid for systems where $q_e\phi_0\ll k_BT_e$ and $T_i\ll T_e$. The new length parameter can therefore be introduced, in accordance with Eqn\eqref{ChiDef1}, as
\begin{equation}
\lambda_\phi=\bfrac{-\en\phi_0}{q_e\sum^K_kZ_kn_{k,\infty}}^{1/2}
    \label{LambaPhiDef1}
\end{equation}
giving the distance to shield a disturbance $\phi_0$ in a plasma medium of k species, corresponding to a general plasma sheath thickness. In the high temperature limit, where the thermal energy of the plasma is comparable to the potential, $\phi_\infty=k_B(T_e+\sum^K_kZ_k^{-1}T_k)/q_e$, applying the quasi neutral identity $q_en_e=\sum_k^Kq_kn_k$, recovers the general Debye length.
\begin{equation}
\lambda_\phi=\bfrac{-\en k_b/q_e^2}{n_e/T_e+\sum^K_kZ_k^2n_{k,\infty}/T_k}^{1/2}=\lambda_D
\label{ShieldingDebyeLengthCompare}
\end{equation}
Using the general shielding length in Eqn.\eqref{LambaPhiDef1}, the physical significance of $\chi$ becomes more obvious, as the ratio between the plasma sheath and the body length for charged flow over a body $\chi=r_0/\lambda_\phi$, so a high $\chi$ represents a relatively large plasma sheath, and therefore a large collection surface for incoming ions. 
\par

\section{Ionospheric Aerodynamics Response Surface}
\begin{figure}[h!]
         \centering
     
         \begin{minipage}[c]{0.95\textwidth}
             \centering
             \includegraphics[width=0.95\textwidth]{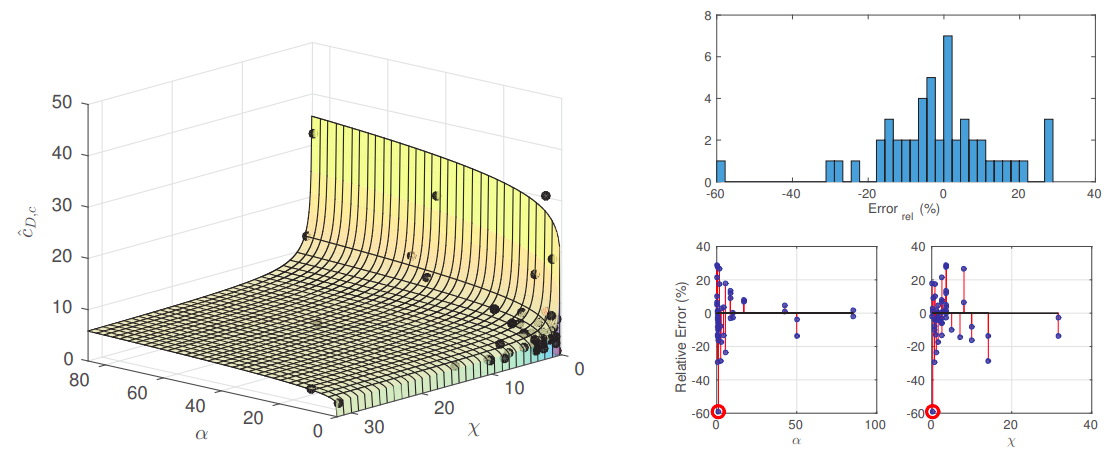}
              \centering
         \end{minipage}
        \captionsetup{width=0.7\linewidth}
         \caption{Reproduced from \cite{CaponBible}, comparison of the response surface specified by Eqn.\eqref{ResponseSurfaceGeneral} with experimental data, showing agreement of within 40$\%$ in all cases save one}
         \label{CC_ResponseSurface}
     \end{figure}

In Capon et al \cite{CaponMain}, a response surface is developed to find drag as a function of $\alpha_k$ and $\chi$. While this response surface is only designed for single species flows, it provides a physics based function to equate wake features to drag coefficients, and therefore an important basis for study of mixed flows.
Findings by Capon et al\cite{Caponthesis} also indicated that the bulk of the variation in drag forces is accounted for by variation in $\alpha_k$ and $\chi$, and proposed a response surface to capture this variation as a function of the form 
\begin{equation}
C_{D,C}=f(\alpha_k)+f(\chi)+f(\alpha_k,\chi)
\label{ResponseSurfaceGeneral}
\end{equation}
Where $C_{D,C}$ was a drag coefficient representing both charged and neutral drag forces.  $f(\alpha_k)$ and $f(\chi)$ are intended to capture variations in $C_{D,C}$ from $\alpha$ and $\chi$ respectively, with the third term accounting for coupled effects. The form of the functions for $\alpha_k$ and $\chi$ are given as
\begin{equation}
f(\alpha_k)=\mathscr{A}(1+2\alpha_k)^{0.5-a},\quad f(\chi)=\mathscr{B}\chi^{-1}
\label{alphaChiForm}
\end{equation}
The term $f(\alpha_k)$ in this has an analogous form to the orbital motion limited impact parameter \cite{AllenOrbitalMotion}. In this case, $\mathscr{A}$ is a geometry dependent constant, while $a$ is to account for the reduction in direct drag caused by ion accelerations. The second term can be made more clear by substituting $\chi=\frac{r_0}{\lambda_\phi}$, where $\lambda_\phi$ is given by Eqn.\ref{LambaPhiDef1} into the Child Langmuir law for sheath thickness ($d_{sh}$) \cite{BenilovLangmuirSheath} around a high voltage cathode, given by
\begin{equation}
d_{sh}=\frac{2^{5/4}}{3}\sqrt{\frac{\en \phi_0}{q_in_i}}=\frac{2^{5/4}}{3}\lambda_\phi=\frac{2^{5/4}r_0}{3}\chi^{-1}
\label{ChildLangmuirSheath}
\end{equation}
thus the term $\chi^{-1}$ is proportional to the ratio of sheath thickness to body size, $d_{sh}/r_0$. In the case of a drag coefficient, this is effectively a measurement of the collection surface of the charged body, roughly analogous to the relationship of incident surface area to $C_D$ in a neutral flow.
The coupling term in Eqn.\ref{alphaChiForm} is given by 
\begin{equation}
f(\alpha_k,\chi)=\mathscr{C}\frac{2\alpha_k^c+\mathscr{D}}{1+\chi}
\label{CouplingTerm}
\end{equation}
where $\mathscr{B},\mathscr{C},\mathscr{D},$ and $c$ are all fitted constants. The response surface was formulated by fitting the constants in Eqns.\ref{alphaChiForm}-\eqref{CouplingTerm} to the drag coefficients of a range of single species, mesothermal plasma flows, giving an empirical response surface (Fig.\ref{CC_ResponseSurface})
This response surface shows good agreement with data of higher $\alpha_k$ and $\chi$ values, but shows greater variation in the range $\alpha_k,\chi<10$. As seen in \cite{CaponBible}, and Chapter\ref{C:wakeStructures}, higher $\alpha_k$ \& $\chi$
approach a simple plasma sheath structure, strongly governed by charged motion and the ratio of kinetic energy to potential gradient laid out in $\alpha_k$, while the coupling term and sheath length term become negligible. For $\alpha_k,\chi<10$, the wake structure has significant contributions from distinct features such as bound jets, which are sensitive to small changes in composition, making prediction of this range difficult.
\par
\subsection{$\Ak$: multi-species deflection parameter}
The most prominent limitation of the study performed in \cite{CaponMain} when applied to mixed flows  lies in the use of the $\alpha_k$ parameter in the response surface. While this study was carried out entirely on single species flows, the $\chi$ parameter is independent of number of species, while $\alpha_k$ refers to only one species at any given time. In order to continue with our investigation of mixed-species flows, it becomes necessary to provide an alternative response surface formulation. In order to develop further the response surface work already done by \cite{CaponBible}, it is ideal that this a formulation is composed of an alternative composition for the parameter $\alpha_k$. In order to be used in place of $\alpha_k$, this parameter (referred to as $\mathbb{A}$ for the present) would need to meet the following criteria:
\begin{enumerate}
    \item $\mathbb{A}$ should still be proportional to the ratio of potential to kinetic energy, while taking into account the contribution of each individual species. 
    \item In the limit where only one species is present in the flow, $\Ak\rightarrow\alpha_{k_0}$ for species $k_0$.
    \item From condition 2, it follows that the value of $\Ak$ for a flow with more than one species must take values in between the values of $\ak$ for each species.
    \item $\Ak$ should be dimensionless.
\end{enumerate}
It follows from these that $\Ak$ must contain terms accounting for variation in not only individual $\ak$ values, and the relative densities of individual ion species. \\
From these conditions, the formulation of $\Ak$ used in this work was given by the sum of each $\ak$, scaled by their respective $\beta_k$ values
\begin{equation}
\Ak=\sum^K_k\beta_k\ak
\label{AlternativeA}
\end{equation}
As $\sum^K_k\beta_k=1$, and each $\beta_k$ acts as a ratio of a species charge density to the total charge density, this results in $\Ak$ fulfilling condition 3. Both components of the product in Eqn.\eqref{AlternativeA} are dimensionless, so fulfill condition 4. As species $k_0$ becomes the dominant source of charge density, $\beta_{k_0}\rightarrow 1$, and all other $\beta_k\rightarrow 0$, fulfilling condition 2. A more intuitive form can be seen by expanding Eqn.\ref{AlternativeA}, and allowing $Z_k=1$ (singly ionised species), and $v_i=v_j,  ( \forall i,j\in K)$ (valid for flows with high ion thermal ratio $S_k$, as with those investigated here) for simplicity gives
\begin{equation}
\Ak=\sum^K_k\frac{n_{k,\infty}}{n_{tot}}\bfrac{q_e\phi_0}{m_kv^2_{k,B}}=\bfrac{q_e\phi_0}{v^2_0n_{tot}}\sum^K_k\bfrac{n_k}{m_k}
\label{ADemo}
\end{equation}
Where $n_{tot}=\sum_k^Kn_k$This shows $\Ak$ as a ratio, still, of potential to kinetic energy, but for any given ion species $i,j$, the value of $\Ak$ lies  between $\alpha_i, \alpha_j$, with the specific ratio of each determined by the relative magnitudes of $n_k/m_k$ for each species.

 \par

\chapter{Plasma flows and wake structures}\label{C:wakeStructures}
\section{Single Species flows}
\label{section:SingleSpeciesflows}
The altitudes comprising low earth orbit contain a mix of plasma species, dominated by singly ionised Hydrogen and Oxygen, with much smaller quantities of singly ionised Helium and Nitrogen (Fig.\ref{IonosphereComposition}). For simplicity, we assume that the dominant factor in any mixed species phenomenon was the charge to mass ratio, while any differences in scattering behaviour or chemical reactions between species were neglected. As the plasmas investigated in this investigation are largely collisionless on the scales being analysed, this assumption is quite intuitive, but is expanded upon further in Section \ref{section:Assumptions}. 
The same mesh (shown in Fig.\ref{BoundaryConditions}), flow velocity, and ion/electron temperatures were used for each case shown in this section (values given in Table \ref{table:Cons}). As investigating the influence of variations in $r_0$, $T_i/T_e$ and $v_{\infty}$, would be computationally infeasible, the analysis focused on variations in $\rho_{k,\infty}$ and $\phi$, and corresponding $\Ak$ and $\chi$.
\par
The structure of the plasma sheath surrounding an object in LEO is a critical component of the drag experienced by the object, with small changes in sheath structure accounting for significant variations in the magnitude and location of drag interactions on a body \cite{CaponBible}. A number of key wake features are analysed in this thesis, most notably the Prandtl-Meyer expansion fan \cite{PrandtlMeyerFan}, bound/unbound ion jets, and the ion void. A representative single species O+ flow is shown in Fig.\ref{SingleFlow}, with flow parameters given in table in appendix \ref{C:InitTable}. The expansion fan front $\mu_1$, ion void, and intersection of two bound jets are marked. These features will be discussed in detail in the following section.
\begin{figure}[h!]
         \centering
     
         \begin{minipage}[c]{0.95\textwidth}
             \centering
             \includegraphics[width=0.99\textwidth]{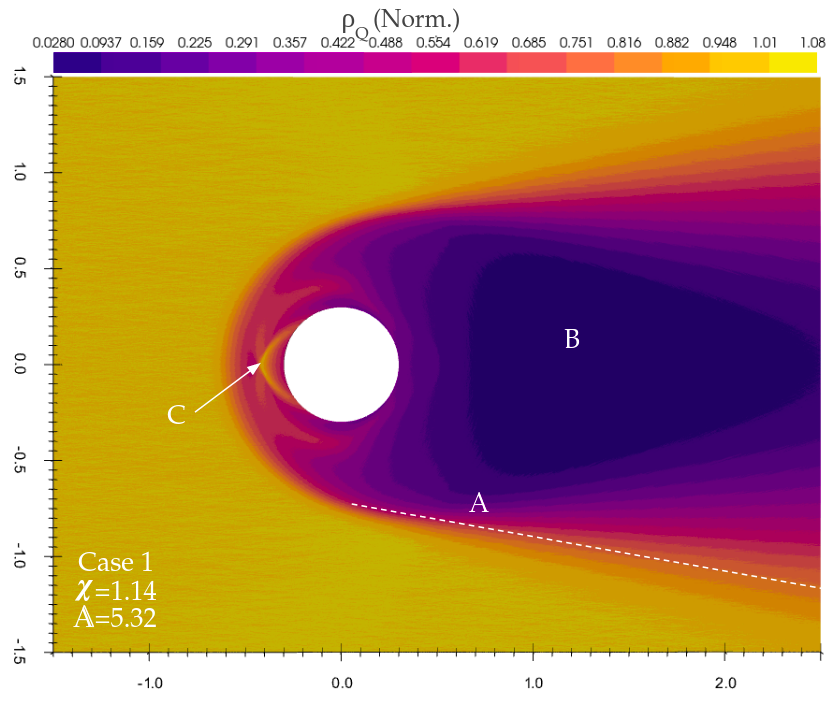}
              \centering
         \end{minipage}
        \captionsetup{width=0.9\linewidth}
         \caption{ Charge density of oxygen plasma flow  over a charged cylinder (Case 1), presented in units of main flow charge density. Key wake features are marked, showing A: expansion fan, B: ion void, and C: Bound jets}
         \label{SingleFlow}
     \end{figure}
    \begin{table}
        \begin{center}

\begin{tabular}{||c | c||} 
 \hline
 Parameter & Value \\ [0.5ex] 
 \hline\hline
 $T_i$ & $1531$ $K$\\ 
 \hline
 $T_e$ & $1997$ $K$\\
 \hline
 $v_{\infty}$  & $7500$  $ms^{-1}$ \\
 \hline
 $Z_k$ & $-1$\\
 \hline
 $r_0$ & 0.3 $m$  \\ [1ex] 
 \hline
\end{tabular}
\end{center}

\caption{Table of constant parameters of interest  (table repeated in Appendix \ref{C:InitTable} }
\label{table:Cons}
\end{table}
\subsection{General sheath and void regions}
\begin{figure}[h!]
         \centering
     
         \begin{minipage}[c]{0.95\textwidth}
             \centering
             \includegraphics[width=0.99\textwidth]{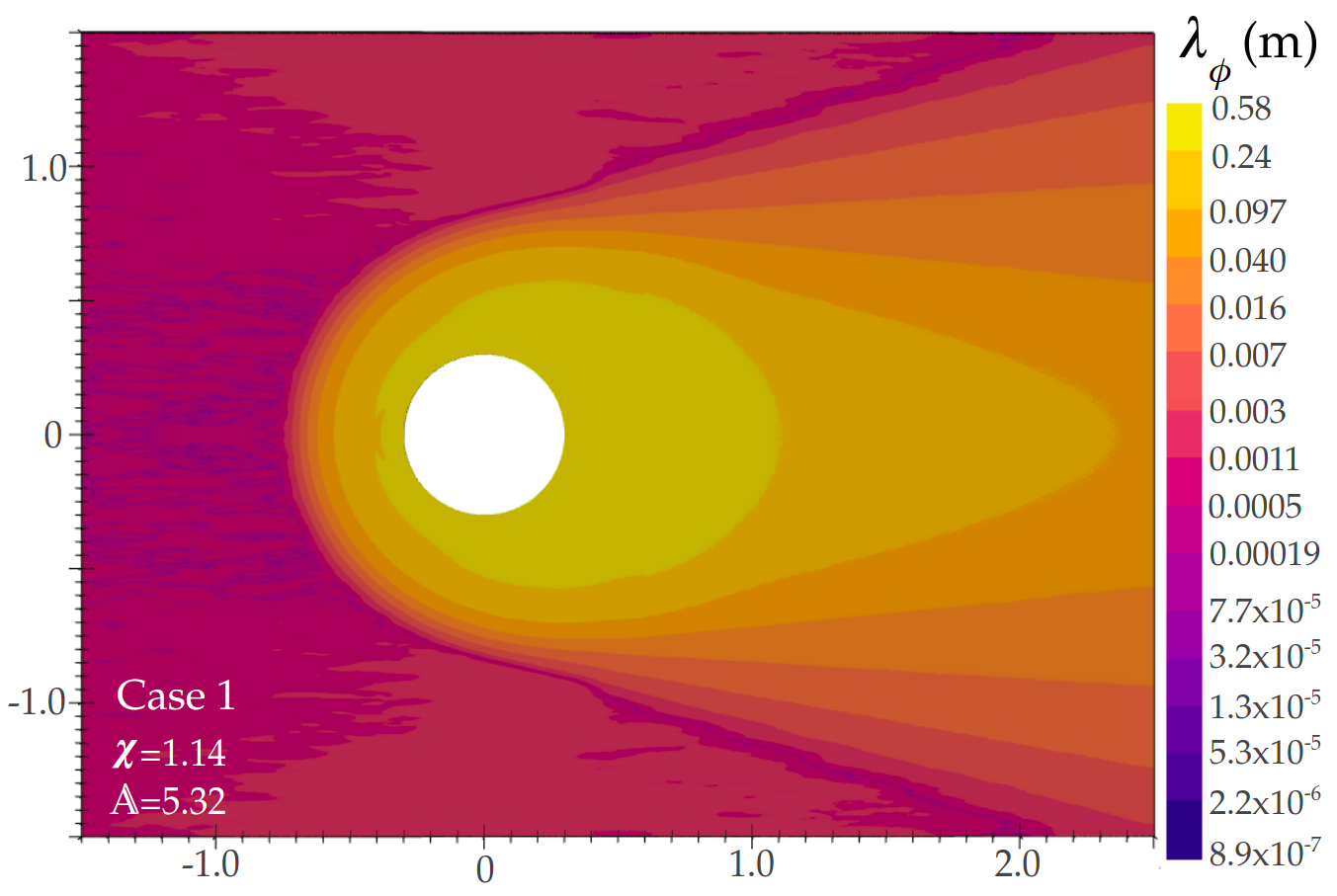}
              \centering
         \end{minipage}
        \captionsetup{width=0.9\linewidth}
         \caption{Plot of local shielding length as defined in Eqn.\ref{ShieldingDebyeLengthCompare} for Case 1, plotted on a log scale. Particles in an area with a low shielding length (such as the bulk flow) are influenced very little by nearby potential disturbances, while forces in regions with a high shielding length (such as the wake void region) are influenced by the charge distribution of a much larger area. }
         \label{ShieldingLength108}
     \end{figure}

The shape of the plasma sheath is notably different from its corresponding neutral flow equivalent. Most notably, the collection surface extends beyond the surface of the object itself, in a visually similar way to a mach shock front. The edge of the plasma sheath in the fore-wake is characterised by a rapid drop in shielding length (Fig.\ref{ShieldingLength108}) as ions enter a highly rarefied region, and are accelerated towards the central potential of the body. Oxygen ions that contact the central potential are neutralised, and not displayed. These neutral particles pass out of the simulation without any collisional interactions, so can be neglected from consideration. As the flow passes around the body, a low density region remains in the wake, forming an ion void. This low density region has a longer shielding length than any other region in the wake, and as a result, ions on the upper and lower boundaries of the ion void are deflected by the central potential, causing the wake to repopulate much more quickly than would be expected by pressure-driven expansion.\\
The ion void is also functionally devoid of electrons, with the high thermal velocity electrons instead remaining in the sheath boundary, resulting in almost no regions sustaining an overall negative charge. \\ 

\subsection{Prandtl-Meyer Expansion Fan}
\begin{figure}[h!]

               \makebox[\textwidth][c]{\includegraphics[width=1.2\textwidth]{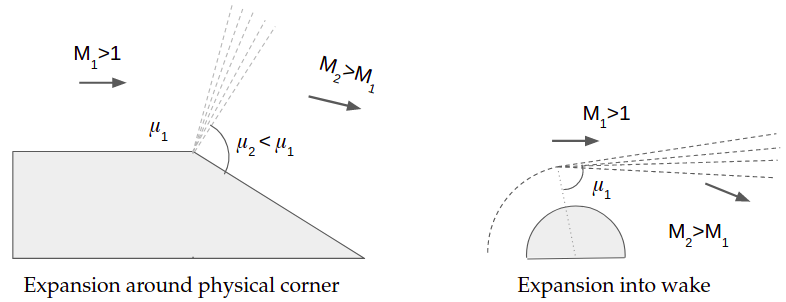}}%
        \captionsetup{width=0.7\linewidth}
         \caption{Prandtl-Mayer expansion fan, showing supersonic flow around a sharp corner (left) and expansion into a wake region, with mach waves in both cases indicated with dashed lines.}
         \label{PrandtleMayerDiagram}
     \end{figure}

A Prandtl-Meyer expansion fan is the result of a supersonic flow expanding into a low pressure region as it rounds a convex corner. It consists of an infinite number of mach waves emitted from a point, forming a 'fan'. In the case of a hard corner, as in Fig.\ref{PrandtleMayerDiagram}, the waves diverge from the corner itself . In the case of a plasma flow over a charged cylinder, the edge of the plasma sheath acts as a corner, forming an expansion fan on the edge of the wake. The angle of the first mach wave $\mu_1$, and final mach wave ($\mu_2$) are given by \begin{equation}
    \mu_1=arcsin(\frac{1}{M_1}),\quad \mu_2=arcsin(\frac{1}{M_2})
    \label{PrandtleMayerAngle}
\end{equation}
where $M_1$ and $M_2$ are the ion acoustic mach numbers of the flow before and after expanding around a corner, respectively \cite{PrandtlMeyerFan}. Using the example of case 1 (Fig.\ref{SingleFlow})(Details in appendix \ref{C:InitTable}), with a mach number of 1.16, the outer mach wave propagates at the expected angle of $\mu_1$. The inner mach waves are influenced not only by pressure driven expansion, but also attraction to the central potential of the satellite body, causing propagation angles greater than $\mu_2$, as well as a curving of the fan structure (seen in Fig.\ref{SingleFlow}).
     
\subsection{Orbital motion limited theory and Bound jets}
For a charged probe in a quasi-neutral plasma, the orbital motion treatment considers the motion of individual ions within the plasma. Positively charged particles from the main flow are deflected towards the negative probe, with the impact parameter $h$ (perpendicular distance from axis through probe), and distance of closest approach related by 
\begin{equation}
    h=r_B\left(1-\frac{\phi_0}{K_i}\right)^{1/2}
    \label{general impact parameter}
\end{equation}
where $K_i$ is the ratio of kinetic energy to charge for ion $i$ (the kinetic energy of an ion is given by $q_eZ_iK_i$). We then consider the case where the impact parameter has a minimum value of $r_c$ for a given $eV_0$, corresponding to an impact parameter denoted $h_*$. This 'critical impact parameter' has a corresponding minimum distance of closest approach $r_c$, at which an incident ion will perform an orbit, then leave the perturbation region formed by the probe \cite{AllenOrbitalMotion}. Any ions with an impact parameter less than $h_*$ have no distance of closest approach, and thus will be accelerated towards the probe and captured. The value for this critical impact parameter for an ion with speed $v_i$ around central potential $\phi_0$ is given by \cite{FORTOVDustyPlasmas} as
\begin{equation}
    h_*=r_0(1-\frac{2q_e\phi_0}{mv_i^2})^{1/2}
    \label{CriticalImpactParameter}
\end{equation}
Considering the specific case of a weakly shielded plasma flow over a charged body in LEO, where the ions flow velocity is much greater than their thermal velocities, the potential field ceases to be cylindrically symmetrical, with a higher potential in the wake due to its low space-charge density.
\par  For a weakly shielded flow ($\chi\gg 1$), the sheath edge lies farther from the body than the critical impact parameter, and the sheath is orbital motion limited (OML). In this case, the ion collection of the sheath is limited by this critical impact parameter, with ions approaching with $h>h_*$ being deflected, and not making contact with the charged body. This sheaths boundary is also defined by this $h_*$ more than by the sheath length equation in Eqn.\ref{ChildLangmuirSheath}.
Flows in which $h_*$ is greater than the distance to the sheath edge from the Langmuir-Child expression (Eqn.\ref{ChildLangmuirSheath}) are referred to as 'sheath limited'.\par
\begin{figure}[h!]
         \centering
             \includegraphics[width=\textwidth]{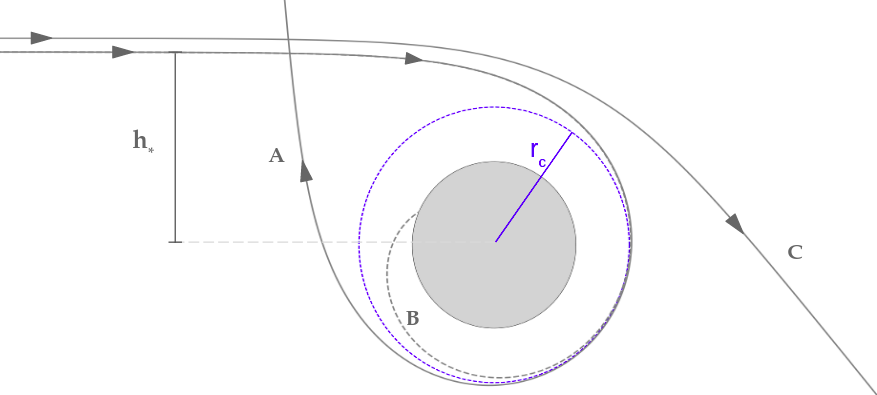}
              \centering
        \captionsetup{width=0.9\linewidth}
         \caption{ Bound jets approaching a charged cylinder at impact parameters close to the critical impact parameter $h_*$, showing radius of closest approach $r_c$, and critical impact parameter $h_*$. Particle paths show effect of a range of impact parameters, A:slightly higher than $h_*$(large deflection, unbound jet), B:less than $h_*$ (bound jet), and C: significantly greater than $h_*$ (small deflection, unbound jet) }
         \label{OMLDiagram}
     \end{figure}
\begin{figure}[h!]
           \makebox[\textwidth][c]{\includegraphics[width=0.9\textwidth]{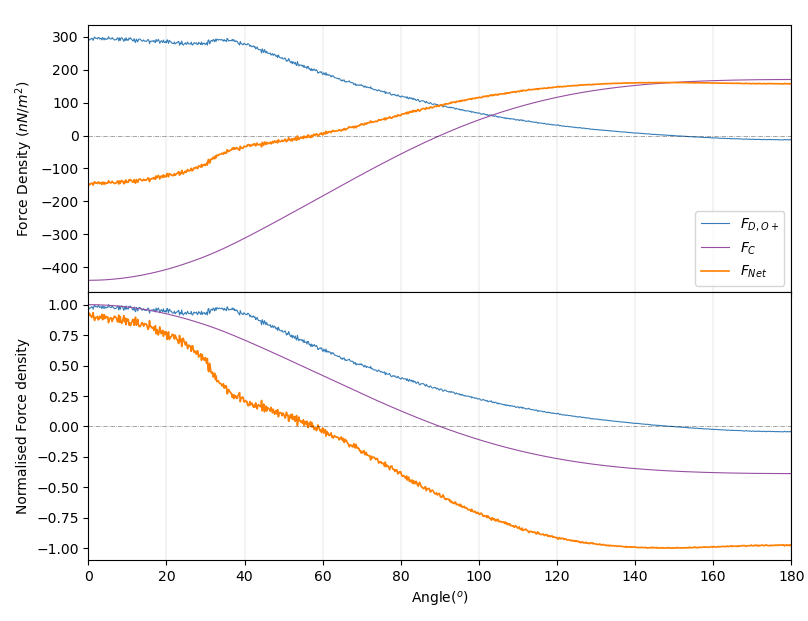}}%

              \centering
        \captionsetup{width=\linewidth}
         \caption{ Plots of force density on cylinder for Case $\#1$ as a function of the angle from direction of travel, where the bottom plot shows components normalised to have the force at $0^{\circ}$ as 1, showing relative magnitudes. $F_{D,O+}$, $F_C$, and $F_{Net}$ are the direct force (from particle collisions), force from electromagnetic interactions, and net force, respectively. The bound jet point of contact is clearly visible as a peak in the direct force around $35\dg$ }
         \label{108Components}
     \end{figure}

In an OML flow, ions entering the sheath at close to the critical impact parameter are pulled into paths towards the main body, that will often pass around in wide arcs (Fig. \ref{OMLDiagram}) before contacting the central potential (Bound jets). In Fig.\ref{SingleFlow}, the bound jet highlighted has passed through over $270^{\circ}$ of rotation before contacting the body. 
Ions with impact parameters greater than $h_*$ will be deflected by the body, or pulled into orbital arcs, but ultimately not contact the body (referred to as deflections, or unbound jets).\\
     \\
To understand the significance of these bound jets on the total drag, we consider an ion following path B in Fig.\ref{OMLDiagram}. This ion will contribute a charged drag in the rear $180^{\circ}$ of its arc, as well as a charged thrust in the portion of its arc spent in front of the satellite. Eventually, the impact on the satellite body will transfer energy comparable to what would be expected from a direct collision with the body from an ion with an impact parameter of 0. The energy exchange imparted by this ion can therefore contribute either a net drag or a net thrust, depending entirely on which point it makes contact with the body surface. These jets, as a collection of such ions, therefore form a significant contribution to the total drag forces on the body, with the actual point of impact of a jet varying greatly with differing wake structures (elaborated further in Section \ref{section:mixedFlows} ). Using case $\#1$ (pure O+ plasma flow)  as an example, plotting the magnitude of force density around the cylinder (Fig.\ref{108Components}) shows an increase in the direct drag at $35\dg$ corresponding to the impact point of the bound jet visible in Fig.\ref{SingleFlow}. Notably, the bound jet does not have any visible peaks in the charged force, as the jet passes around almost $360\dg$ of arc before making contact, contributing evenly to electromagnetic forces across this range. Due to the collisionless nature of this flow, this can cause two overlapping flows of different directions in a number of regions.

\newpage
     \newpage
\subsection{Effect of $\Ak$ and $\chi$  on wake structures}
\begin{figure}[h!]
         \centering

             \includegraphics[height=0.76\textheight]{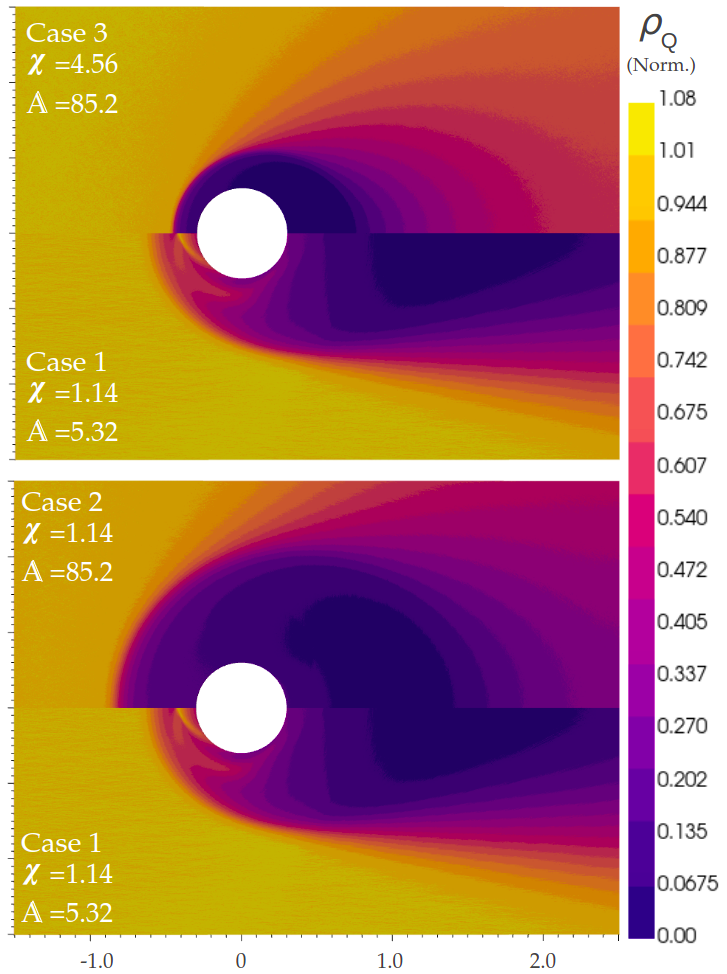}
              \centering
        \captionsetup{width=\linewidth}
         \caption{Plots of cases 1, 2,and 3 as a function of relative ion number density of plasma (normalised to a main flow density of 1$m^{-3}$). Case 1 is entirely O+ plasma, while cases 2 and 3 are entirely H+ plasma, with either the same charge density (Case 2) or Mass density (Case 3) as Case 1. Full parameters are given in table in Appendix \ref{C:InitTable}. Direction of flow is from left to right.}
         \label{Single Species flows}
     \end{figure}

\begin{figure}[h!]
         \centering
             \includegraphics[width=0.9\textwidth]{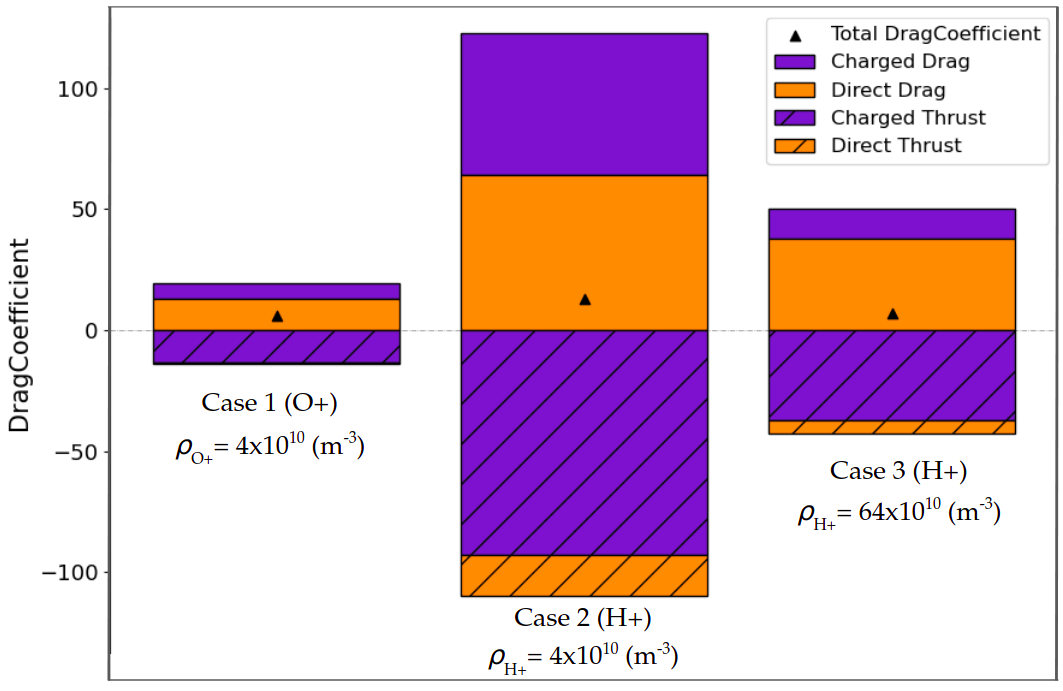}
              \centering
        \captionsetup{width=0.8\linewidth}
         \caption{Relative drag coefficient of three single species cases in Fig\ref{Single Species flows}, with charged and direct drags, thrusts, and net $C_d$ marked. Notably, direct forces contribute significantly to thrusts in cases 2 and 3 (sheath limited H+), but not in case 1 (pure oxygen), where the flow is orbital motion limited.}
         \label{SingleSpeciesBars}
     \end{figure}

To highlight the effects of varying $\Ak$ and $\chi$ on the various wake structures, we introduce flows $\#2$ and $\#3$ (Fig.\ref{Single Species flows}), H+ flows with the same flow parameters as flow 1, with the exception that case 2 and 3 are  flows of Hydrogen plasma with either the same charge density (case 2) or mass density (case 3) as case 1 (full parameters in table in appendix \ref{C:InitTable}). \\
Most notable in these flows is the way the lower charge to mass ratio of hydrogen causes it to fill the wake region more quickly than O+ plasma. The critical impact parameter predicted by OML theory for these flows is also 3.9m, compared to the much lower 1.0 of the O+ flow, therefore we do not expect to see significant presence of bound jets, though the tight wake structure allows for a significant contribution of direct thrust from ions pulled into short orbital paths (Fig.\ref{SingleSpeciesBars}). These plots also serve to illustrate general trends within single species flows, characterised by changing $\Ak$ and $\chi$.\\ For lower values of $\chi$, the sheath size grows (as in cases 2 and 3), as the ability of the flow to shield from the influence of $\phi_0$ is diminished. For highly shielded flows (high $\chi$), the ion collection is dominated by direct drag, with sheath enhanced ion collection being minimal. As $\chi\rightarrow 0$, the sheath expands, and direct drag forces increase to an asymptote. The increase in effective collection area by the expanded sheath contributes the increased direct drag, with the asymptote occuring as the flow becomes orbital motion limited (case 1, both case 2 and 3 are sheath limited). As the sheath thickness increases, the indirect wake drag also increases, while the direct thrust forces on the rear surface of the body increases up to the orbital motion ion collection limit.

\newpage
 Ions continue to be deflected past the OML sheath thickness, but are no longer collected, and do not contribute to direct thrusts, only contributing a net charged drag. Therefore as $\chi \rightarrow 0$, we expect to see both direct drag and direct thrusts increase up to an OML asymptote, while charged drags and thrusts continue increasing. As $\chi \rightarrow \infty$, the flow becomes highly shielded, and the flow approaches a neutral, high density flow.\\
Using $\Ak$ in the single species limit $\ak$, we can see more clearly its influence over sheath structures. Most notably, we see
in Eqn.\ref{CriticalImpactParameter} that $\ak$ governs the critical impact parameter, and therefore the OML limited behaviour of weakly shielded flows. As $\ak$ rises, we expect the critical impact parameter to also rise, so the wake approaches sheath limited behaviour. For kinetically dominated flows with a low $\ak$, the sheath structure is kinetically dominated, and the sheath boundary is close to the body. Providing the flow is not sheath limited, the outer edge of the wake is estimated by the critical impact parameter $h_*$ (Eqn.\ref{CriticalImpactParameter}). \par
\begin{figure}[h!]
         \centering
             \includegraphics[width=0.85\textwidth]{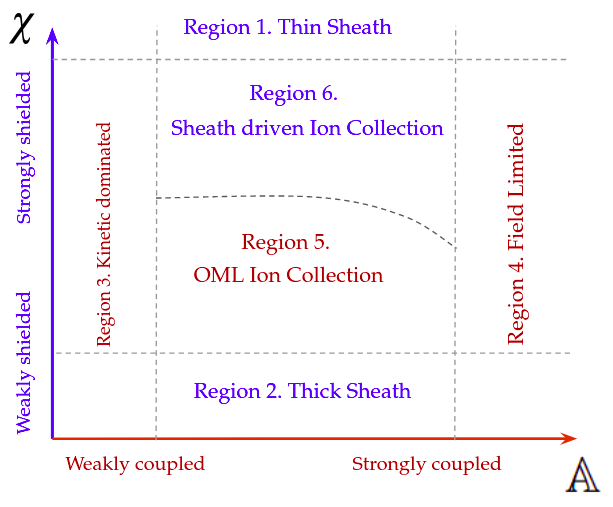}
              \centering
        \captionsetup{width=\linewidth}
         \caption{Limiting behaviour of sheaths with two main scaling parameters, showing areas dominated by the ion deflection parameter $\Ak$ in red, and areas dominated by the shielding ratio $\chi$ in blue. Scale is arbitrary and the exact location of these boundaries is a subject for further study.}
         \label{RegimeDiagram}
     \end{figure}

In summary, the sheath structure of a single species plasma flow over a charged body in LEO conditions is governed mostly by two parameters, $\Ak$ and $\chi$. Fig.\ref{RegimeDiagram} summarises the regimes of a mesothermal plasma flow with varying $\Ak$ and $\chi$. In limiting cases: As $\Ak\rightarrow 0$, the sheath approaches a kinetically dominated structure much like a neutral flow. Similarly, as $\chi\rightarrow\infty$, the sheath becomes thin, with ion collection only marginally higher than in a neutral flow.
As $\Ak\rightarrow\infty$, the sheath structure is dominated by field effects, with sheath driven ion collection and a sheath boundary defined by the equation for sheath radius around a stationary probe given in \cite{BenilovLangmuirSheath} Eqn.\eqref{ChildLangmuirSheath}. In between these extreme limits of sheath behaviour, the ion collection in the sheath is dominated by either the Langmuir sheath edge, or the critical impact parameter of the flow, leading to either sheath limited ion collection or OML ion collection, with the boundary determined by the particular $\Ak$ and $\chi$ of the flow. The present study focuses on the regime in Regions 5 and 6 of Fig.\ref{RegimeDiagram}, where the Langmuir sheath size is comparable to the critical impact parameter, and both play a significant role in sheath structure.

\section{Mixed Species flows}\label{section:mixedFlows}
In order to investigate the effects of mixed species interactions on charged aerodynamics, a range of cases were prepared. These cases focused on mixtures of  singly ionised H+ and O+ flows, falling into two broad categories:
\begin{itemize}
    \item Constant Mass: cases kept to a mass of $64\eten \frac{amu}{m^3}$, with varying ratios of O+ and H+, representing sheath limited (high $\chi$) wake structures.
    \item Constant Charge: cases kept to a charge density of $4\eten \frac{q_e}{m^3}$, with varying ratios of O+ and H+. These cases are typically OML, with defined bound jets.
\end{itemize}
 The full details of each case can be found in the table in Appendix \ref{C:InitTable}

\subsection{Sheath limited mixed flows}
\begin{figure}[h!]
         \centering
             \includegraphics[width=\linewidth]{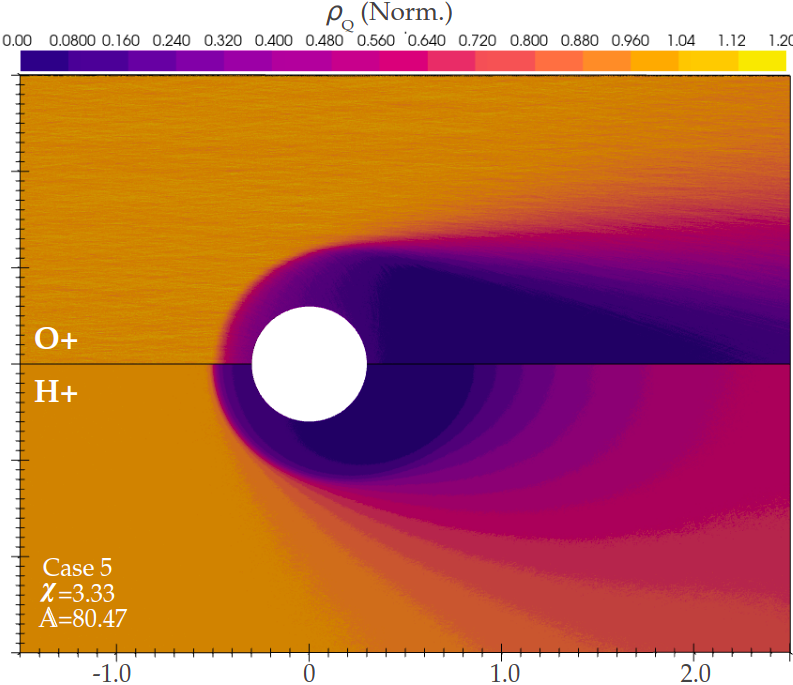}
              \centering
        \captionsetup{width=\linewidth}
         \caption{Ion charge density plot of H+ and O+ in case 5, normalised to a freestream mass density of 1 ($\frac{kg}{m^3}$). case 5 is a sheath dominated flow, and both O+ and H+ share the same sheath front on the forward face.}
         \label{SheathLimitedContours}
     \end{figure}

In mixed, collisionless flows, the only interaction between heavy and light ion species is assumed to be electrostatic, so it is useful to consider the different behaviours of the two component species, as they can form distinct, overlapping wake structures. In Fig.~\ref{SheathLimitedContours} , a flow of equal parts H+ and O+ by mass is presented (Case 5), showing the differing wake structures of the two component species. \\ 
 Ions approaching the wake edge from the main flow experience little to no force from the body potential until reaching the main wake edge, where the sudden decrease in density allows them to be accelerated towards the satellite. As a result, the oxygen component to the sheath in Fig.\ref{Components94}, which would normally be limited by $h_*$, is limited instead by the boundary of the H+ sheath, which shields the O+ from experiencing any coulomb force outside this region. As such, both the H+ and O+ components of the flow are sheath limited, and share a $d_{sh}$ value, despite drastically different $\ak$ parameters for each. The H+ can be seen to fill the wake void much more quickly than O+, shielding the expansion fan structure from the potential, and largely removing the curved fan structure shown in O+ in Fig.\ref{SingleFlow}. As a result of this shielding, almost all of the direct thrust is the result of H+ being pulled from the wake structure.
 \begin{figure}[h!]
         \makebox[\textwidth][c]{\includegraphics[width=\textwidth]{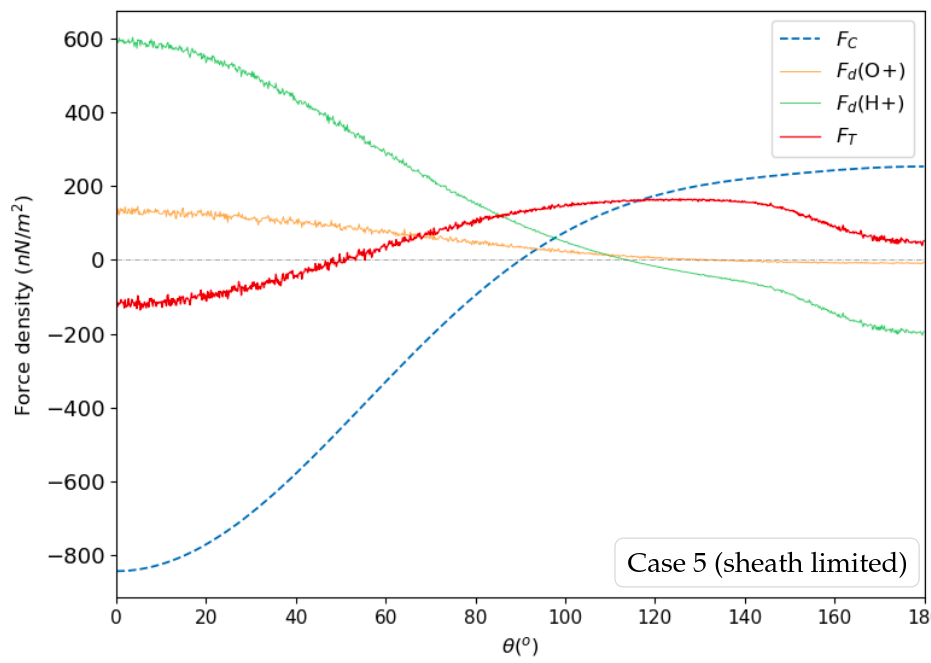}}
        \captionsetup{width=0.9\linewidth}
         \caption{Force density on the cylinder (case 5) as a function of angle from flow direction. $F_{D}$, $F_C$, and $F_{T}$ are the direct force (from particle collisions), force from electromagnetic interactions, and net force, respectively.}
         \label{94ForcebyAngleComponents}
     \end{figure}

Despite both species contributing equal mass density, it is also apparent that H+ contributes more to direct drag (see Fig.\ref{94ForcebyAngleComponents}, as lighter H+ ions are accelerated into the surface from much greater impact parameters, causing an effectively larger collection area for H+, while O+ ion collection is largely limited to ions with impact parameters comparable to the radius of the body  (Fig\ref{94ForcebyAngleComponents}). \par

\begin{figure}[h!]
  \centering
        \makebox[\textwidth][c]{\includegraphics[width=0.8\textwidth]{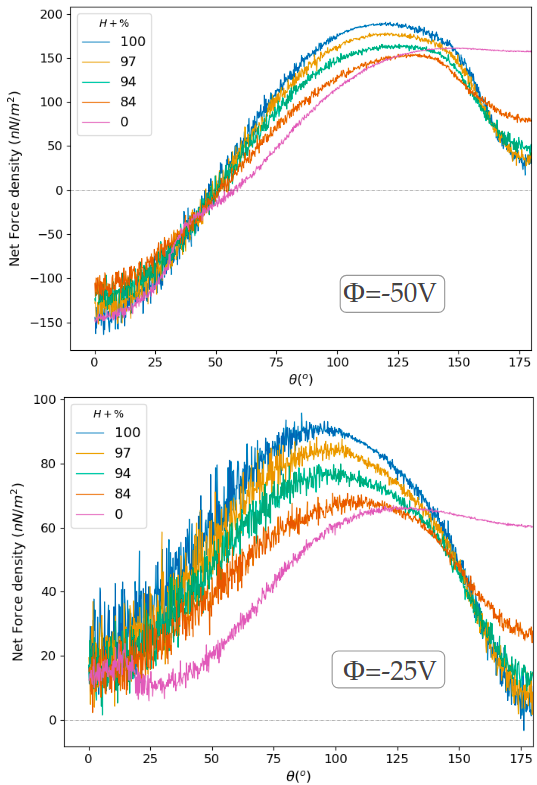}}
              \centering
        \caption{Net Force as a function of angle from flow direction for sheath limited cases with a constant flow mass density shared between cases. Cases are a mix of O+ and H+, with percentage H+ (by mass) displayed.  Two plots shown differ by central potential $\Phi$.}
        \label{ConsMassForce}
\end{figure}

\begin{figure}[h!]
         \centering
             \includegraphics[width=0.8\linewidth]{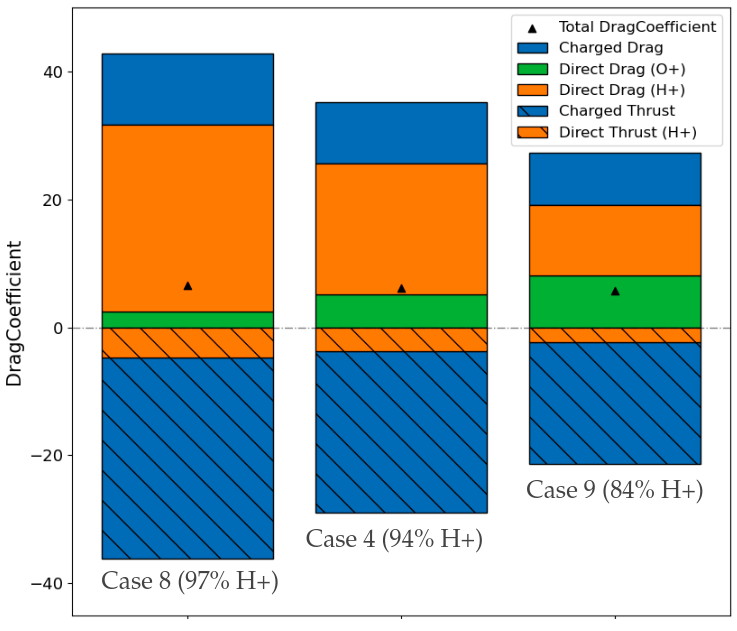}z
              \centering
        \captionsetup{width=\linewidth}
         \caption{Relative drag coefficient contribution of the direct and charged components of 3 mixed flows with a constant mass density, with O+ concentration increasing from left to right. O+ direct thrust accounted for less than 1$\%$ of total thrust, so was neglected from plot. H+$\%$ is in terms of charge density.}
         \label{ConsMassBarPlot}
     \end{figure}

As the ratio of $\rho_{H+}:\rho_{O+}$ is varied from entirely O+ to entirely the lighter H+, the effect of the higher charge density can be seen in the peak in net drag around 100$\dg$ (Fig.~\ref{ConsMassForce}). This peak corresponds to a region in which small angle deflections of ions from the flow into the wake contribute a charged drag, while ions impact at shallow angles themselves, resulting in a region experiencing no thrust contribution (and thus contributing most of the drag on the body).\\
Due to the small angle deflections of uncaptured ions contributing to charged thrusts and drags without corresponding direct forces, the net force on the front surface of the high potential cases (Fig.\ref{ConsMassForce}) is actually a thrust force , while the forces on the rear face contribute the drag (note this is still a net drag). This is only true for the high potential case, as the low potential case experiences a drag over its entire surface, peaking at around $90\dg$.\\
Also noteworthy is the fact that the force on the cylinder in the range $160\dg-180\dg$ varies by almost 50$\%$ between mixed flows (Fig.\ref{ConsMassBarPlot}), despite the H+ concentration (which contributes majority of direct forces in these flows (Fig.\ref{ConsMassBarPlot})) varying by less than 20$\%$. One reason for this is the fact that H+ ions tend to be pulled into short paths to the body instead of being deflected into the wake, hence their contribution to the charged forces is closely matched by an enhanced direct force from the coulomb acceleration. As a result, the O+ ions contribute a large portion of the unbalanced forces through ion deflections. This drag is therefore increasing with raising O+ concentration, which raises by over 100$\%$ between mixed flows in Fig. \ref{ConsMassForce}

\newpage
\subsection{Orbital Motion Limited Mixed flows}
\begin{figure}[h!]
         \centering
             \includegraphics[width=\linewidth]{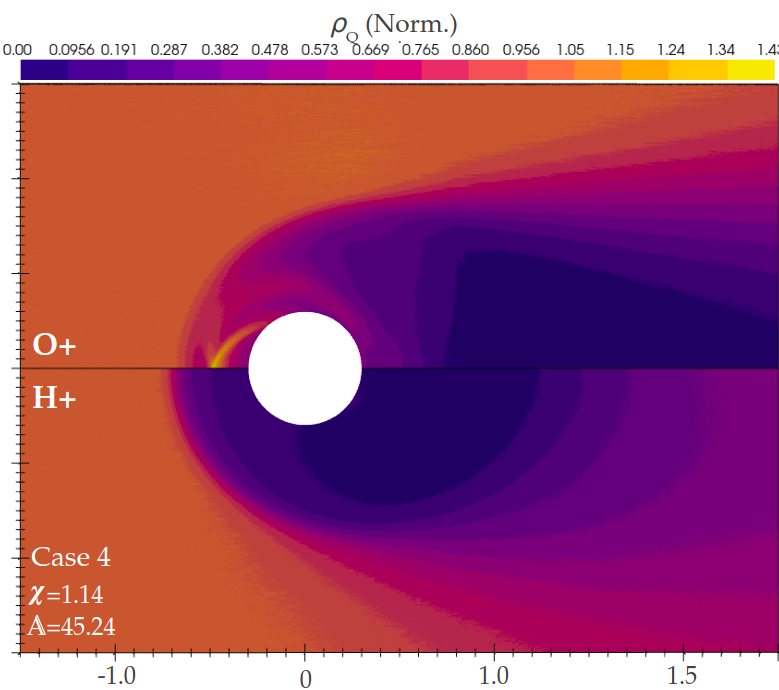}
              \centering
        \captionsetup{width=\linewidth}
         \caption{Ion charge density plot of O+ (top) and H+ in case 4, presented in units of main flow density. case 4 is an OML dominated flow,  with strong bound jets visible in the O+ distribution.}
         \label{Components94}
     \end{figure}

\begin{figure}[h!]
   \makebox[\textwidth][c]{\includegraphics[width=0.89\textwidth]{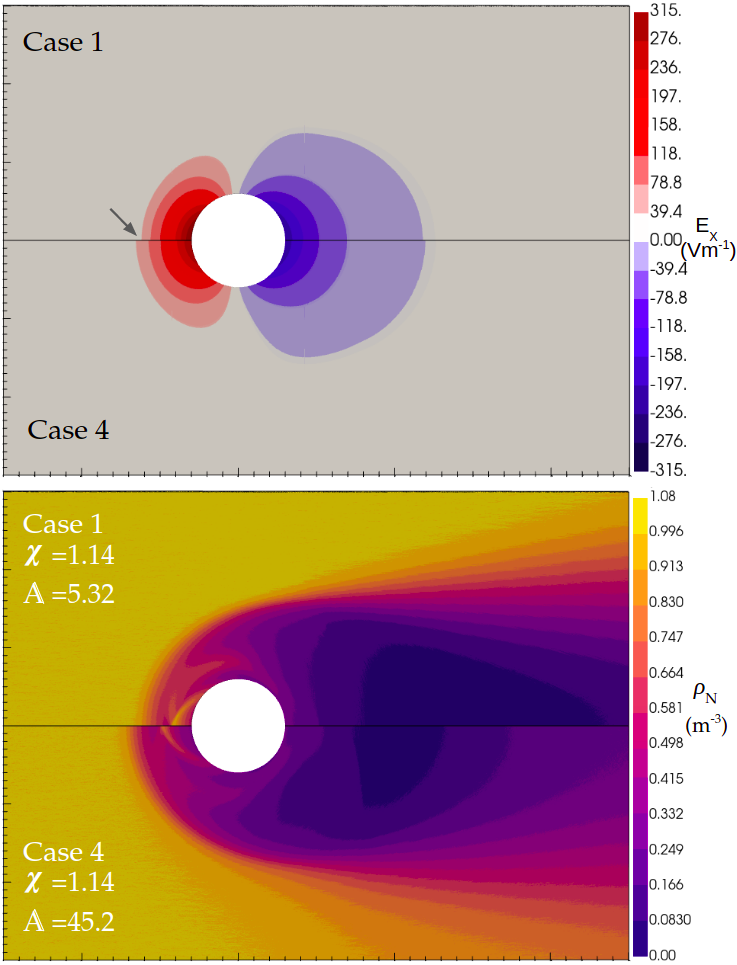}}

        \caption{Plots of the horizontal component of electric field (Top), and total ion number density for cases 1 and 4. Discontinuity in forebody $E_x$ is marked, showcasing higher electrostatic acceleration on approaching O+ , which leads to the higher critical impact parameter $h_*$ observed in case 4. Ion density presented in units of main flow density.}
        \label{RaisedOMLEvidence}
\end{figure}

In case 4 (Shown in Fig.\ref{Components94}), the mass density of H+ was greatly reduced compared to O+. As a result,the sheath edge is defined by the orbital motion limited critical impact parameter $h_*$ of the O+ contribution, rather than the langmuir-Child sheath, so the H+ sheath boundary moves inwards to the edge of the O+ sheath. As in the sheath limited case above, given the same sheath edge in the fore-body region, H+ fills the wake more quickly, similarly 'straightening' the curved fan structure seen in the pure O+ flow, by shielding the fan edge from the influence of the charged body.\\
\\Notably, the actual sheath width has raised in the mixed flow compared to the pure O+ flow, despite both sheaths being defined by the OML collection of the O+ ions. This behaviour is unintuitive in this case, as the critical impact parameter for O+ predicted in Eqn.\ref{CriticalImpactParameter} is unchanged in both flows, and therefore this behaviour is a result of an interaction between the two ion species. Specifically, as the H+ and O+ enter the wake, the lighter H+ ions are accelerated more quickly towards the cylinder (as can be seen in the slightly broader H+ sheath), lowering the charge density and shielding of ions in this region. Conversely in the wake of the body, H+ ions fill the ion void much more quickly than the corresponding O+ ions, resulting in a highly shielded region with reduced electric field. 
The net effect of these two alterations to the potential field experienced by O+ is a field in which forebody ion acceleration is increased, while the wake potential remains largely unchanged (Fig.\ref{RaisedOMLEvidence}), resulting in the heavier O+ ions effectively entering the sheath with a higher velocity. This additional horizontal acceleration leads to a raised $h_*$, and associated broader sheath than in the pure O+ flow. While $\chi$ and $\alpha_{O+}$ are unchanged between cases 1 and 4, the variation is qualitatively accounted for in the raising $\Ak$ parameter.
Note that in Fig.\ref{RaisedOMLEvidence}, the electric field and subsequent acceleration of ions has been presented as a proxy for the ion velocity, in order to demonstrate the increased O+ velocity at the sheath boundary. The velocity plot was not presented, as overlapping streams of particles lead to average velocity visualisations to be quite misleading, showing regions of overlapping high velocity flows as the sum of components instead. The velocity of particles at the boundary was also sampled directly to confirm this enhanced O+ velocity.
\newpage

\begin{figure}[h!]
   \makebox[\textwidth][c]{\includegraphics[width=0.75\textwidth]{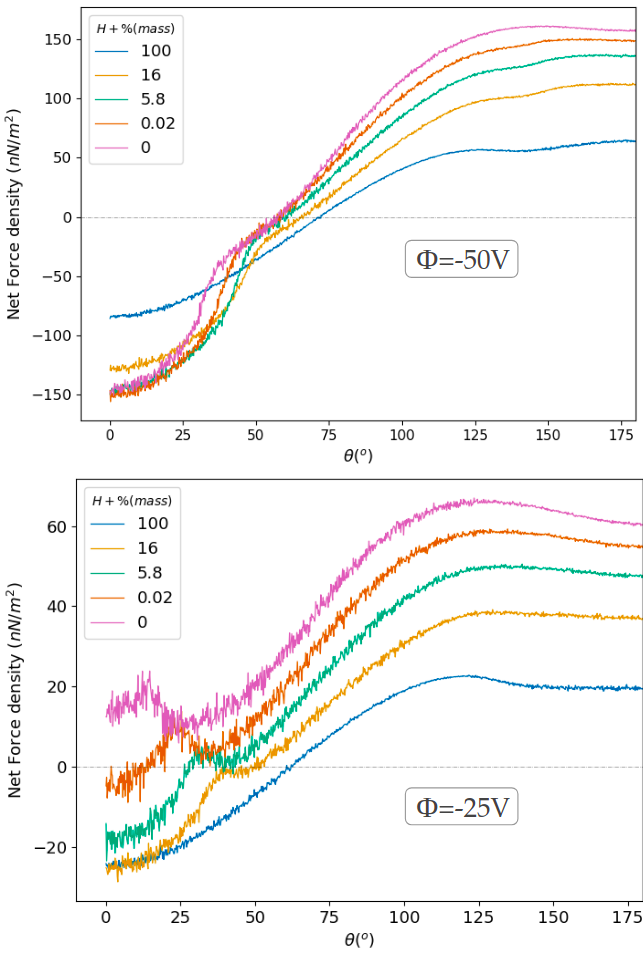}}
        \caption{Net Force as a function of angle from flow direction for OML cases with a constant flow charge density shared between cases. Two figures differ by ratio of H+ to O+ ions, with the percentage H+ (by mass density) marked on plot. }
        \label{ConsChargeForce}
\end{figure}

Figure \ref{ConsChargeForce} presents net force acting on an object for different fraction of charged oxygen to charge hydrogen for different angles around the cylinder. The bound jets in OML limited sheaths represent a significant component of the drag forces experienced, and the angle of impact of these jets is therefore an important parameter to consider in this analysis. In Fig.\ref{ConsChargeForce}, the jet corresponding to case 4's higher impact parameter makes contact at a shallower angle of incidence, having travelled a longer path around the body. This longer path has the effect of increasing the charged thrust experienced by the body, while the shallower angle of incidence lowers the amount of momentum transferred in the direction of motion, having the net effect of lowering the O+ contribution to the drag forces compared to the pure oxygen case. This pattern continues with increasing concentrations of H+ (Fig.\ref{ConsChargeForce}), causing the bound jet impact point to shift around the cylinder (visible as a peak in the $30\dg -60\dg$ range in direct force). \\
It is worth considering that this trend does not necessarily represent a reduction in drag in all cases, rather the shifting impact point corresponds to a reduction in drag in this case only because this jet was incident at a very shallow angle in the case used here. For a different initial velocity and impact point, the increased H+ could easily result in a raised drag force contribution from the bound jet.



\newpage



\chapter{Summary of work and Discussion}\label{C:con}

With an increasingly congested LEO environment comes a necessity of effective orbit propagation techniques, allowing for prediction and management of satellite orbits. Historically, many orbit propagation techniques used the drag equation (Eqn.\ref{DragEquation}), with a drag coefficient of 2.2, and the assumption that neutral aerodynamic effects dominate most LEO conditions. The increasing use of high power electronics in LEO, coupled with a greater understanding of ionospheric composition, calls some of the fundamental assumptions in this method into question, and shows the necessity of a physics based method for evaluating ionospheric aerodynamics.\\
Work done at UNSW Canberra by Capon et al created a hybrid DSMC-PIC solver, and used this to develop an empirical response surface to predict ionospheric drags on  satellites, incorporating both charged and neutral drag effects. This response surface relied on a set of dimensionless parameters describing the plasma-body interaction, specifically a shielding parameter $\chi$ and ion deflection parameter $\ak$. This response surface relied on assumptions of single species dominated flows, which comprise the majority of LEO, with the exception of a band of 200km height.\\
This work aimed to expand the ionospheric aerodynamic modelling work of Capon et al to a form capable of accounting for mixed species interactions, and analyse the interactions of mixed species in plasma flows, in order to extend the ionospheric aerodynamics model of Capon et al.\\
PdFoam was extended, and tested, to record instantaneous charged force values on the cylinder. This code was used in performing a numerical convergence and optimisation study for multiple simulation parameters, in order to minimise computational cost while still providing reliable results. The analysis of mixed species flows focused on a series of simulations of flows over a uniformly charged cylinder, with differing ratios of O+ to H+ ions.\par

The parameter $\Ak$ was introduced as a mixed species alternative to the ion deflection parameter $\ak$, chosen for the property of being representative of the average ion deflection parameter for the flow, and being a linear combination of existing pi-groups. The aerodynamic forces on a range of mixed flows were analysed. The key findings are summarised here: 
\par
Orbital motion limited flows with high O+ experienced an increase in sheath size with the introduction of small concentrations of H+. O+ Ions are also seen to enter the sheath with increased velocities over those seen in pure O+ flows. This increased velocity and sheath size corresponds to a greater critical impact parameter, and a significant change in the location of impact of the bound O+ jets. Bound jet impact locations were seen to have an average ion flux of roughly double the regions around them, making the location of impact an important parameter in determining the total drag acting on the satellite.\\
Introduction of small quantities of H+ into the O+ dominated flow also increased the total drag coefficient of the flow, which can be largely explained by three factors:
\begin{itemize}
    \item Increased sheath size represents a greater ion collection area, and more forebody drag.
    \item Increased O+ velocity entering sheath results in greater forebody momentum exchange, and less ions collected on rear face, lowering direct  O+ thrust while increasing direct drag.
    \item H+ ions contribute the same charge density as the equivalent concentration of O+ ions, with a lower mass. This mean any electromagnetic momentum exchange from ion deflection will result in a higher drag coefficient for a flow with greater relative H+ concentration, as the drag coefficient is based on the ratio of incident KE density to total force.
\end{itemize}

Introducing small concentrations of O+ into sheath limited flows with a high H+ concentration relative to O+ showed negligible change in the sheath structure, which remained limited by the H+ sheath boundary (although the lowering charge density imposed by replacing H+ with equal masses of O+ results in broadening of the sheath as expected from Eqn.\eqref{ChildLangmuirSheath}). Despite a similar sheath structure, the drag properties change notably if the ratio of O+:H+ concentration is varied, as O+ ion collection is largely limited by the thin sheath. This leads to negligible direct thrust from O+, but a significant contribution (roughly twice that which would be expected from a neutral oxygen flow) to direct forebody drag. H+ contributed similarly to direct forebody drag, but the lighter mass enables a direct thrust as well. In this way, introducing O+ into a sheath limited, H+ dominated flow resulted in very little change to the charged aerodynamics, with most of the variation in Fig.\ref{ConsMassForce} being explained by the varying space-charge density.\\
These results indicate that the most significant area for charged Aerodynamic effects is regimes dominated by heavy ions, where the effects of small concentrations of lighter ions are more prominent. This is consistent with experimental findings of \cite{MixedSheathKnewIt}, where Ar+ ions were found to move at a greatly increased velocity across the plasma sheath in a plasma flow with small concentrations of the lighter He+. This also represents a significant deviation from drag predictions in the regimes where $\frac{\rho_{O+}}{\rho_{H+}}\approx 10$  (750-900km in Fig.\ref{IonosphereComposition}).\par
The parameter $\Ak$ is shown to describe the qualitative behaviour of the flows where it concerns the critical impact parameter and sheath thickness, but falls short in quantitative predictions, and use of the $\Ak$ parameter in Capon et als response surface model as a substitute for $\ak$ results in overestimation of drag by 25-70$\%$ in all mixed cases (Shown in table in Appendix.\ref{C:InitTable}). One likely cause of this discrepancy is the interference between sheaths with different limiting behaviour; As each set of $\chi$ and $\ak$ describe a flow, and its ion collection behaviour, the combination of the two flows, described by one $\chi$ and $\Ak$, may not take into account the fact that the species with the wider sheath will be blocked from its own shielding behaviour by the species with the narrower sheath. As an example, in Case 4 (Fig.\ref{Components94}), the H+ flow on its own would be expected to form a sheath with width 3.9m (from Eqn.\ref{ChildLangmuirSheath}), but is instead shielded from electromagnetic effects until it reaches the surface of the OML O+ sheath. Therefore, the H+ ion collection is significantly different to what is described by $\chi$ and $\alpha_{H+}$ for this system, a discrepancy not taken into account by the parameters used in $\Ak$.
\\
For the purposes of quantitatively predicting drag on mixed species, the method of introducing a substitute parameter for $\ak$ falls short, as the drag experienced is a function of both the individual ion deflection parameters of the species, and inter-species interactions that vary in their intensity between OML and sheath limited flows. This implies that if this method is to be pursued, it is important to analyse the validity of the coupling term in Eqn.\eqref{ResponseSurfaceGeneral}, stated by Capon et al to be the subject of potential future work. Ideally, this term would be altered to accurately reflect the limiting behaviours in mixed species flows, and the tendency of the sheath structure to be dominated by whichever species would naturally result in a thinner sheath. The establishment of such a method is the subject for further research.\\

\newpage
\section{Review of assumptions made}
\label{section:Assumptions}
A number of assumptions were employed in this study, many of which were also employed and analysed in the defining work performed by Capon et al \cite{Caponthesis}. A summary of these assumptions, and associated assessment of their validity is presented here.\\
\begin{itemize}
\item \textbf{2d flows are assumed to be representative of 3d flow aerodynamics over a cylinder:} The fundamental mechanics of the plasma sheath are unchanged between 2 and 3 dimensions, and the dimensionless parameters derived in Chapter \ref{C:DimensionlessParameters} are derived in a general case, and hold for any number of dimensions. In a 3d flow, we would expect to see two major discrepancies from the wake structures shown here, namely the dispersal of bound jets into less tightly collimated structures from a less uniform central potential well(and subsequent reduction of drag coefficient), alongside an increased collection area on the ends of the cylinder, where the ions can be collected from the z axis (out of plane in the 2d simulations) as well. The exact magnitude of the discrepancies between 2d and 3d is therefore a function of the relative magnitude of these two effects. In \cite{CCExperiment}, 2d and 3d simulations of this kind are shown to produce similar predictions over the potentials simulated here, with increasing discrepancies with raising potential. There is no indication that mixed species aerodynamics would alter this interaction significantly.
\item \textbf{Bulk flow velocity of different ion species is similar between species} These simulations results are based on the assumption that the bulk flow of an ionospheric plasma in LEO has the same velocity for each species present, excluding thermal deviations. This is currently supported by ionospheric measurements of particle velocity, showing H+ and O+ kinetic energy to be distributed according to average flow speed and temperature only \cite{IonSpeedMeasurements}. 
\item \textbf{Flow conditions used are  representative of LEO mixed flows} In this work, H+ dominated cases are taken to represent sheath limited wake flows, while O+ dominated cases are taken to represent OML wake flows. The effects of heavy O+ on an OML H+ dominated sheath or H+ on a sheath dominated O+ flow are not investigated here. However, in the regions in which mixed flows are seen to be most prevalent (700-1000km), the satellite orbit necessitates a speed high enough for predominantly O+ flows to always be OML, while the areas in which the flow is significantly fast for H+ dominated flows to be orbital motion limited are at a much lower altitude ($\approx 300km$), at which both neutral drag is dominant, and H+ concentration is negligible. Therefore it is reasonable to model H+ dominated mixed flows as sheath limited in LEO, and O+ dominated mixed flows as orbital motion limited.
\item \textbf{Approximation of a satellite in LEO as a perfect cylinder} While this assumption obviously neglects many of the complexities of actual satellite geometries, it provides an effective model for general aerodynamic studies, and the cylindrical symmetry allows for analysis of general charged aerodynamics without considering the contribution of angle of attack, lift forces or similar asymmetrical effects. The greatest source of uncertainty in applying this method to prediction of a satellites motion is likely to therefore result from the actual satellite geometry, as a long, thin shape moving perpendicular to its long axis is going to experience a greater charged force than a more compact shape with the same surface area, due to the increased collection area.
\item \textbf{Approximation of satellite as uniformly charged and insulating} As with the cylindrical assumption, we know a satellite is unlikely to be uniformly charged if it is made of insulating materials, and charge asymmetries are likely to arise from a combination of any high power equipment used on the satellite, and the asymmetrical nature of bound jet currents. The effects of these charge asymmetries, much like body geometry asymmetry, is therefore very specific to the satellite under analysis. Notably, for flows with a low $\chi$ parameter, any charge assymetries are unlikely to significantly affect the general wake behaviour, which will still be dominated by the net charge of the body for ions entering the sheath (as the sheath boundary occurs at a distance much greater than the distances over which the charge asymmetry occurs, so the satellite potential approaches that of a point charge)
\item \textbf{Magnetism is assumed to play no part in Ionospheric aerodynamics} At the altitudes considered, the gyral radius of ions is of significantly larger size than most bodies (with the notable exception of the ISS) \cite{Caponthesis}, and therefore ions can be approximated as moving in straight lines for distances considered in this work.
\end{itemize}

\subsection{Further Research Avenues}
\begin{itemize}
\item \textbf{Gas surface interactions} The work in Moe et al \cite{MoeMoe} provides evidence that the drag coefficient over a neutral satellite can vary by as much as 30$\%$ depending on the particle-wall interaction model in use. In the context of charged aerodynamics, this affects the ratio of electromagnetic to neutral momentum exchange for a particle accelerated into the satellite body, and has significant potential to alter the drag coefficient. In a collisionless plasma (such as those simulated for this thesis), we do not expect the gas-surface interaction to alter the wake formations significantly, so the net effect would be one of scaling the direct force, rather than altering the fundamental findings of this work. The gas-surface interaction is largely affected by the level of atomic oxygen, with high concentrations of oxygen causing a layer to build up on the satellite surface, making these gas interactions diffuse. This effect is notably only likely to be significant in the O+ dominated cases here, as atomic oxygen largely ceases to exist in the atmosphere above 700km, where what little oxygen is present is directly photo-ionised.
\item \textbf{Bound Jet dynamics} Bound jet impact points in the case of cylinders are seen to represent both a significant contribution to the force in the area, as well as a significant discontinuity in the ion current (40$\%$ increase in force and $60\%$ increase in ion current (Fig.\ref{108Components}). The location of this bound jet is very sensitive to the potential distribution, as well as the velocity of ions entering the sheath, therefore quite sensitive to the relative ion concentrations in a mixed flow. Of particular interest in the analysis of bound jet forces is the potential torque provided to an asymmetric satellite by these bound jets, with jets often exceeding main flow kinetic energy by 20$\%$, in perpendicular directions.
\item \textbf{Differential charging} This analysis focuses on the charge held by a perfectly conducting body, assumed to maintain  constant charge throughout the course of it orbit. In reality, the conditions leading to satellite charging result in a varying charge with local plasma conditions, which will then affect the interaction of a plasma with its environment. Numerous studies have been performed on the general case of a satellite charging in LEO, often using experimental data \cite{ISSCharging}\cite{OlsenPurvisCharging}\cite{FERGUSONCharging}, but none has yet been performed incorporating differential charging into ionospheric aerodynamic forces for nonconductive, or low conductivity bodies. Of particular interest for this work is the effect of the differential charging on bound jets, and the unbalanced ion current provided to the region of impact.
\end{itemize}




\appendix{}
\chapter{Table of Initial conditions}\label{C:InitTable}

  Table of values (next page) for key parameters used in simulations referenced in Chapter~\ref{C:wakeStructures}. All parameters refer to initial conditions for main flow, not necessarily final state. First table shows parameters kept constant across all species and cases, second table shows variable parameters by case.
\begin{table}
\centering
\begin{tabular}{||c | c||} 
 \hline
 Parameter & Value \\ [0.5ex] 
 \hline\hline
 $T_i$ & $1531$ $K$\\ 
 \hline
 $T_e$ & $1997$ $K$\\
 \hline
 $v_{\infty}$  & $7500$  $ms^{-1}$ \\
 \hline
 $Z_k$ & $-1$\\
 \hline
 $r_0$ & 0.3 $m$  \\ [1ex] 
 \hline
\end{tabular}
\end{table} 

\begin{adjustbox}{center}


\begin{tabular}{||c||c|c|c||c|c||c|c||c| c||} 
 \hline
 Case & $\rho_{N,O+} \; (m^{-3})$ & $\rho_{N,H+} \; (m^{-3})$ & $\phi_B (V)$ & $\chi$ & $\Ak$ & $C_D$ & $C_C$ & $C_{Tot}$&$C_{\frac{Tot}{Pred}}$\\ [0.5ex] 
 \hline\hline
 1 & $4\ex{10}$ & $0$ & -50 & 1.14& 5.32 &12.86 &-7.14 &5.72&0.775\\ 
 
  2  & $0$ & $4\ex{10}$ & -50 & 1.14& 85.16 &47.37 &-34.6 &12.77&1.02\\ 
   3 & $0$ & $64\eten$ & -50 & 4.56& 85.16 &31.99 &-25.12 &6.88&0.868\\ 

  4  & $2\eten$ & $2\eten$ & -50 & 1.14& 45.24 &15.65 &-9.16 &6.49&0.579\\
  
  5 & $2\eten$ & $32\eten$ & -50 & 3.33& 80.47 &21.9 &-15.91 &5.99&0.696\\
    6 & $1\ex{10}$ & $3\ex{10}$ & -50 & 1.14& 65.2 &19.54 &-12.05 &7.49&0.62\\ 

  7 & $3\eten$ & $1\eten$ & -50 & 1.14& 25.28 &13.97 &-7.94 &6.03&0.59\\

  8 & $1\eten$ & $48\eten$ & -50 & 3.99& 83.53 &27.0 &-20.57 &6.43&0.78\\

  9 & $3\eten$ & $16\eten$ & -50 & 2.49& 72.56 &16.76 &-11.03 &5.72&0.617\\ 

  10 & $4\ex{10}$ & $0$ & -25 & 1.61& 2.66 &7.86 &-3.30 &4.56&0.825\\ 
 
  11  & $0$ & $4\ex{10}$ & -25 & 1.61& 42.6 &29.08 &-19.19 &9.89&1.01\\ 
 
  12 & $1\ex{10}$ & $3\ex{10}$ & -25 & 1.61& 32.6 &12.03 &-6.0 &6.02&0.64\\ 
  13  & $2\eten$ & $2\eten$ & -25 & 1.61& 22.62 &9.56 &-4.44 &5.12&0.585\\ 
 
  14 & $3\eten$ & $1\eten$ & -25 & 1.61& 12.64 &8.52 &-3.72 &4.8&0.61\\ 
 
  15 & $0$ & $64\eten$ & -25 & 6.45& 42.6 &20.96 &-15.07 &5.89&0.91\\ 
 
  16 & $1\eten$ & $3\eten$ & -25 & 41.77 & 41.77 &17.63 &-12.25 &5.38&0.806\\ 
 
  17 & $2\eten$ & $32\eten$ & -25 & 4.7& 40.23 &14.32 &-9.33 &4.99&0.719\\ 
 
  18 & $3\eten$ & $16\eten$ & -25 & 3.52& 36.28 &10.89 &-6.22 &4.67&0.632\\ 

 \hline \hline

\end{tabular}
\end{adjustbox}
  \\ \\ \\
  Columns are, from left to right: Case number, O+ number density, H+ number density, body potential, Shielding ratio, mixed-species deflection parameter, direct drag coefficient, charged drag coefficient, total drag coefficient, and ratio of net drag coefficient to predicted coefficient (using Eqn.\eqref{ResponseSurfaceGeneral}).\\

\chapter{Symbols and Notation Guide}\label{C:NotationGuide}
Laid out here is the general guide for meaning of variables and symbols used in this thesis
\begin{description}
   \item \textbf{List of symbols used and their meanings} 
   \item[$\phi$] Electric potential
   \item[$n_k, \rho_{N,k}$] number density of ion species k
   \item[$C_D$] drag coefficient
   \item[$r_0$] Characteristic length scale (Satellite radius)
   \item[$\ak$] Ion deflection parameter
   \item[$\beta_k$] Ion coupling parameter
      \item[$\lambda_D$] Debye length
      \item[$\lambda_\phi$] general shielding length
      \item[$S_k$] Ion thermal ratio
      \item[$\mu_k$] electron energy coefficient
      \item[$\chi$] General body shielding ratio
      \item[$Z_k$] ionisation level of species k
      \item[$q_k$] Charge of species k (in Coulombs)
      \item[$M_k$] Ionic mach number of species k
      \item[$f_k$] phase space distribution function of species k
      \item[$\Bar{T}$]  Maxwell stress tensor
      \item[$S$] Poynting vector
      \item[SX] Mesh scaling parameter
      \item[$N_{Equiv}$] Macroparticle number scaling parameter
      \item[$\Delta t$] algorithm timestep
      \item[$\rho_Q$] Charge density (in Coulombs)

    \item \textbf{Subscripts meanings}
    \item[0] A quantity with subscript $_0$ pertains to the satellite body (In dimensional analysis, this also pertains to characteristic dimensions)
    \item[k] pertaining to an ion species k
    \item[$\infty$] a value taken an arbitrary distance away from the body (in the main flow)
\end{description}
\bibliographystyle{unsrt}

\bibliography{BiblioNew.bib}

\begin{thebibliography}{10}

\bibitem{CaponBible}
L.~Brown.
\newblock Rarefied plasma aerodynamics for leo objects in the ionosphere.
\newblock {\em Air Force office of scientific research, Australia}, 2018.

\bibitem{MixedSheathKnewIt}
G.D. Severn, Xu~Wang, Eunsuk Ko, N.~Hershkowitz, M.M. Turner, and
  R.~McWilliams.
\newblock Ion flow and sheath physics studies in multiple ion species plasmas
  using diode laser based laser-induced fluorescence.
\newblock {\em Thin Solid Films}, 506-507:674--678, 2006.

\bibitem{ThreeSpeciesPlasmaSheath}
M.~Hatami and I.~Kourakis.
\newblock Characteristics of plasma sheath in multi-component plasmas with
  three-ion species.
\newblock {\em Scientific Reports}, 12, 04 2022.

\bibitem{UCSSatellites}
{UCS Satellite Database}.
\newblock \url{https://www.ucsusa.org/resources/satellite-database}.
\newblock Accessed: 06/06/2021.

\bibitem{AlJazeera}
Spacex satellites’ near-misses fuel us-china tensions.
\newblock
  \url{https://www.aljazeera.com/economy/2021/12/29/bbspacex-satellites-near-misses-with-lab-fuel-us-/china-tensions}.
\newblock Acessed: 13/03/2022.

\bibitem{debrismitigationRex}
D.~Rex and P.~Eichler.
\newblock The possible long term overcrowding of leo and necessity and
  effectiveness of debris mitigation measures.
\newblock {\em Proceedings of the First European Conference on Space Debris,
  Darmstadt, Germany}, 1993.

\bibitem{Kessler}
D.~Kessler and B.~Cour-Palais.
\newblock Collision frequency of artificial satellites: The creation of a
  debris belt.
\newblock {\em Journal of Geophysical Research: Space Physics},
  83(A6):2637--2646, 1978.

\bibitem{satelliteManeuvering}
D.~Mishne and E.~Edlerman.
\newblock Collision-avoidance maneuver of satellites using drag and solar
  radiation pressure.
\newblock {\em Journal of Guidance, Control, and Dynamics}, 40(5):1191--1205,
  2017.

\bibitem{FuenteDrag}
A.~de~la Fuente.
\newblock {\em Enhanced Modelling of LAGEOS Non-Gravitational Perturbations}.
\newblock PhD thesis, Delft University of Technology, 2007.

\bibitem{MoeMoe}
K.~Moe and M.~Moe.
\newblock Gas–surface interactions and satellite drag coefficients.
\newblock {\em Planetary and Space Science}, 53:793--801, 07 2005.

\bibitem{DoornbosDensityModel}
E.~Doornbos and H.~Klinkrad.
\newblock Modelling of space weather effects on satellite drag.
\newblock {\em Advances in Space Research}, 37(6):1229--1239, 2006.

\bibitem{MSISE00}
J.~M. Picone, A.~E. Hedin, D.~P. Drob, and A.~C. Aikin.
\newblock Nrlmsise-00 empirical model of the atmosphere: Statistical
  comparisons and scientific issues.
\newblock {\em Journal of Geophysical Research: Space Physics}, 107(A12):SIA
  15--1--SIA 15--16, 2002.

\bibitem{WingTheoryAbbot}
I.~Abbott and A.~von Doenhoff.
\newblock Theory of wing sections.
\newblock 1959.

\bibitem{CookDragCoefficient}
G.~E. Cook.
\newblock On the accuracy of measured values of upper-atmosphere density.
\newblock {\em Journal of Geophysical Research (1896-1977)}, 70(13):3247--3248,
  1965.

\bibitem{ISSCharging}
P.~C. Anderson.
\newblock Characteristics of spacecraft charging in low earth orbit.
\newblock {\em Journal of Geophysical Research: Space Physics}, 117(A7), 2012.

\bibitem{Hastings30V}
D.~E. Hastings.
\newblock A review of plasma interactions with spacecraft in low earth orbit.
\newblock {\em Journal of Geophysical Research: Space Physics},
  100(A8):14457--14483, 1995.

\bibitem{OlivieraWeather}
D.~Oliveira and E.~Zesta.
\newblock Satellite orbital drag during magnetic storms.
\newblock {\em pre-print}, 2019.

\bibitem{OlsenPurvisCharging}
R.~Olsen and C.~Purvis.
\newblock Observation of charging dynamics.
\newblock {\em Journal of Geophysical Research}, 88:5657--5667, 07 1983.

\bibitem{ValladoAndFinkleman}
D.~Vallado and D.~Finkleman.
\newblock A critical assessment of satellite drag and atmospheric density
  modeling.
\newblock {\em Acta Astronautica}, 95, 02 2014.

\bibitem{Caponthesis}
C.~Capon.
\newblock {\em Ionospheric Aerodynamics in Low Earth Orbit}.
\newblock PhD thesis, 09 2017.

\bibitem{BenilovLangmuirSheath}
M~S Benilov.
\newblock The child{\textendash}langmuir law and analytical theory of
  collisionless to collision-dominated sheaths.
\newblock {\em Plasma Sources Science and Technology}, 18(1):014005, nov 2008.

\bibitem{CCScaling}
C.~Capon, M.~Brown, and R.~Boyce.
\newblock Scaling of plasma-body interactions in low earth orbit.
\newblock {\em Physics of Plasmas}, 24:042901, 04 2017.

\bibitem{BrundinChargedDrag}
C~L Brundin.
\newblock Effects of charged particles on the motion of an earth satellite.
\newblock {\em AIAA (Am. Inst. Aeron. Astronaut.) J.}, Vol: 1, 11 1963.

\bibitem{Whipple81}
E.~C. {Whipple}.
\newblock {Potentials of surfaces in space}.
\newblock {\em Reports on Progress in Physics}, 44(11):1197--1250, November
  1981.

\bibitem{GODDLaFramboiseCurrentCollection}
R.~Godd and J.G. Laframboise.
\newblock Total current to cylindrical collectors in collisionless plasma flow.
\newblock {\em Planetary and Space Science}, 31(3):275--283, 1983.

\bibitem{Stone1981SpaceAerodynamics}
N.~H. Stone.
\newblock The aerodynamics of bodies in a rarefied ionized gas with
  applications to spacecraft environmental dynamics.
\newblock 1981.

\bibitem{CaponMain}
C.~Capon, M.~Brown, and R.~Boyce.
\newblock Direct and indirect charged aerodynamic mechanisms in the ionosphere.
\newblock {\em Advances in Space Research}, 62, 06 2018.

\bibitem{FORTOVDustyPlasmas}
V.E. Fortov, A.V. Ivlev, S.A. Khrapak, A.G. Khrapak, and G.E. Morfill.
\newblock Complex (dusty) plasmas: Current status, open issues, perspectives.
\newblock {\em Physics Reports}, 421(1):1--103, 2005.

\bibitem{BirdBookDSMC}
G.A. Bird.
\newblock {\em The DSMC method}.
\newblock Createspace Independent Publishing, 2013.

\bibitem{LiuMeanFreePath}
V.~C. Liu.
\newblock Ionospheric gas dynamics of satellites and diagnostic probes.
\newblock {\em Space Science Reviews}, 9, 1969.

\bibitem{CAPONPdFoam}
C.~Capon.
\newblock pdfoam: A pic-dsmc code for near-earth plasma-body interactions.
\newblock {\em Computers \& Fluids}, 149:160--171, 2017.

\bibitem{maxwelStressTensor}
F.P. Miller, A.F. Vandome, and J.~McBrewster.
\newblock {\em Maxwell Stress Tensor}.
\newblock Alphascript Publishing, 2010.

\bibitem{WernerTimestepConstraints}
G.~R. Werner, T.~G. Jenkins, A.~M. Chap, and J.~R. Cary.
\newblock Speeding up simulations by slowing down particles: Speed-limited
  particle-in-cell simulation.
\newblock {\em Physics of Plasmas}, 25(12):123512, 2018.

\bibitem{TimeStabilityConditions}
M.~Vass, P.~Palla, and P.~Hartmann.
\newblock Revisiting the numerical stability/accuracy conditions of explicit
  pic/mcc simulations of low-temperature gas discharges.
\newblock {\em Plasma Sources Science and Technology}, 2022.

\bibitem{BoltzmanEFLuid}
J.~Boerner and I.~Boyd.
\newblock Numerical simulation of probe measurements in a non-equilibrium
  plasma.
\newblock {\em 36th AIAA Plasmadynamics and Lasers Conference}, 2009.

\bibitem{ControlSurfaceAllen}
J.~E. Allen.
\newblock On the drag on an object immersed in a flowing plasma: the control
  surface approach.
\newblock {\em Journal of Plasma Physics}, 73(5), 2007.

\bibitem{GriffithsElectrodynamics}
D.~G. Griffiths.
\newblock {\em Introduction to electrodynamics}.
\newblock Cambridge University Press, 1981.

\bibitem{ConvergenceConditionsThompson}
D.~B. Thompson.
\newblock {\em Numerical Methods 101 - Convergence of Numerical Models}.
\newblock American Society of Civil Engineers, Baltimore, Maryland, 1992.

\bibitem{FEMScaling}
I.~Farmaga, P.~Shmigelskyi, Piotr Spiewak, and L.~Ciupinski.
\newblock Evaluation of computational complexity of finite element analysis.
\newblock pages 213 -- 214, 03 2011.

\bibitem{HallPlasmaVehicleInteraction}
D.~F. Hall, R.~F. Kemp, and J.~M. Sellen.
\newblock Plasma-vehicle interaction in a plasma stream.
\newblock {\em AIAA Journal}, 2(6):1032--1039, 1964.

\bibitem{BeiserScalingLaws}
A.~{Beiser} and B.~{Raab}.
\newblock {Hydromagnetic and Plasma Scaling Laws}.
\newblock {\em Physics of Fluids}, 4(2):177--181, February 1961.

\bibitem{knechtel1964experimental}
E.~D. Knechtel and W.~C. Pitts.
\newblock Experimental investigation of electric drag on satellites.
\newblock {\em AIAA Journal}, 2(6):1148--1151, 1964.

\bibitem{PlasmaExpansionIntoVacuumSamir}
U.~Samir, K.~H. Wright~Jr., and N.~H. Stone.
\newblock The expansion of a plasma into a vacuum: Basic phenomena and
  processes and applications to space plasma physics.
\newblock {\em Reviews of Geophysics}, 21(7):1631--1646, 1983.

\bibitem{Lacina_1971}
J~Lacina.
\newblock Similarity rules in plasma physics.
\newblock {\em Plasma Physics}, 13(4):303--312, apr 1971.

\bibitem{ThermalVel}
C.~M. {Xu}, Y.~Y. {Chen}, R.~J. {Yu}, and Y.~Y. {Zhang}.
\newblock {Influence of particle velocity on the conductivity of dusty plasma}.
\newblock {\em Indian Journal of Physics}, 92(6):799--811, June 2018.

\bibitem{Banks2004ThermalVelOxy}
B.~Banks, S.~Miller, and K.~de~Groh.
\newblock Low earth orbital atomic oxygen interactions with materials.
\newblock {\em NASA TM-220042213233}, 2, 08 2004.

\bibitem{AllenOrbitalMotion}
J~E Allen.
\newblock Probe theory - the orbital motion approach.
\newblock {\em Physica Scripta}, 45(5):497--503, may 1992.

\bibitem{PrandtlMeyerFan}
T.~Meyer.
\newblock {\em Über zweidimensionale Bewegungsvorgänge in einem Gas, das mit
  Überschallgeschwindigkeit strömt}.
\newblock PhD thesis, 01 1908.

\bibitem{CCExperiment}
P.~Lorrain, C.~Capon, R.~Boyce, C.~Maldonado, and A.~Ketsdever.
\newblock Experimental investigation of ionospheric aerodynamics effects.
\newblock {\em AIP Conference Proceedings}, 2132:110003, 08 2019.

\bibitem{IonSpeedMeasurements}
R.~A. Heelis, R.~A. Stoneback, M.~D. Perdue, M.~D. Depew, and W.~A. Morgan~et
  al.
\newblock Ion velocity measurements for the ionospheric connections explorer.
\newblock {\em Space science reviews}, 212(1-2):615–629, 2017.

\bibitem{FERGUSONCharging}
D.~C. Ferguson.
\newblock Chapter 15 - extreme space weather spacecraft surface charging and
  arcing effects.
\newblock In {\em Extreme Events in Geospace}, pages 401--418. Elsevier, 2018.

\end{thebibliography}

\end{document}